\documentclass[12pt,a4paper,final]{iopart}

\usepackage{iopams}  
\usepackage{graphicx}
\usepackage{amssymb}
\usepackage{bm}
\usepackage{color}
\usepackage{amsthm}
\usepackage[disable]{todonotes}

\theoremstyle{definition}
\newtheorem{remark}{Remark}

\usepackage[breaklinks=true,colorlinks=true,linkcolor=blue,urlcolor=blue,citecolor=blue]{hyperref}

\begin{document}

\title[On dimension reduction in Gaussian filters]{On dimension reduction in Gaussian filters}

\author{Antti Solonen$^{1,2,3}$, Tiangang Cui$^{2}$, Janne Hakkarainen$^{4}$, and Youssef Marzouk$^{2}$}
\address{$^1$School of Engineering Science, Lappeenranta University of Technology, Lappeenranta, Finland}
\address{$^2$Department of Aeronautics and Astronautics, Massachusetts Institute of Technology, Cambridge, MA 02139}
\address{$^3$Eniram Ltd, Helsinki, Finland}
\address{$^4$Earth Observation Unit, Finnish Meteorological Institute, Helsinki, Finland}
\ead{\mailto{antti.solonen@gmail.com}, \mailto{tcui001@gmail.com}, \mailto{janne.hakkarainen@fmi.fi},  \mailto{ymarz@mit.edu}}

\begin{abstract}
\textit{A priori} dimension reduction is a widely adopted technique for reducing the computational complexity of stationary inverse problems. In this setting, the solution of an inverse problem is parameterized by a low-dimensional basis that is often obtained from the truncated Karhunen-Lo\`{e}ve expansion of the prior distribution. For high-dimensional inverse problems equipped with smoothing priors, this technique can lead to drastic reductions in parameter dimension and significant computational savings.

In this paper, we extend the concept of \textit{a priori} dimension reduction to non-stationary inverse problems, in which the goal is to sequentially infer the state of a dynamical system. Our approach proceeds in an \textit{offline-online} fashion. We first identify a low-dimensional subspace in the state space before solving the inverse problem (the offline phase), using either the method of ``snapshots'' or regularized covariance estimation. Then this subspace is used to reduce the computational complexity of various filtering algorithms---including the Kalman filter, extended Kalman filter, and ensemble Kalman filter---within a novel subspace-constrained Bayesian prediction-and-update procedure (the online phase). We demonstrate the performance of our new dimension reduction approach on various numerical examples. In some test cases, our approach reduces the dimensionality of the original problem by orders of magnitude and yields up to two orders of magnitude in computational savings.

\end{abstract}

\vspace{2pc}
\submitto{\IP}

\section{Introduction}

An inverse problem converts noisy, incomplete, and possibly indirect observations to characterizations of some unknown parameters or states of a physical system.
These unknowns are often functions defined on a spatial domain and linked to the observables via a forward model. 
High dimensionality due to the numerical discretization of the unknown functions is often viewed as one of the grand challenges in designing scalable inference methods.
This challenge motivates the development of dimension reduction approaches, which exploit the possibly low-dimensional intrinsic structure of the inverse problem, to alleviate the effect of the ``curse of dimensionality.''

%
%
%

A typical inverse problem is ill-posed; the unknowns are not uniquely identified by the observations. This is a joint effect of noisy, incomplete observations and the smoothing properties of the forward model. 
In the Bayesian inference framework \cite{tarantola05, kaipio05}, ill-posedness is addressed by employing a suitable prior distribution and characterizing the posterior distribution of the unknowns conditioned on the observations.
In this setting, the priors often encode structural information about the unknowns, such as spatial smoothness properties. 
This \textit{a priori} structural information opens up the possibility of prior-based dimension reduction, especially in cases where the variation of the high-dimensional unknowns can be explained by a small number of basis functions. 
For instance, the truncated Karhunen-Lo\`eve expansion \cite{karhunen47, loeve78} of the prior distribution is employed in \cite{marzouk09} for identifying such a \textit{a priori} low-dimensional basis in \textit{static} inverse problems.
Computational cost can be greatly reduced by projecting the original high-dimensional unknowns onto the subspace spanned by the resulting low-dimensional basis. 
%


In this paper, we extend this concept of \textit{a priori} dimension reduction to non-stationary inverse problems, in which the goal is to sequentially infer the state of a dynamical system.
Such problems can be solved efficiently using filtering methods, where the posterior prediction from the previous time step is used as the prior for the current state, and the new posterior is obtained by conditioning the current state on data observed at the current time. 
%
The computational difficulty of applying filtering methods to high-dimensional problems stems both from propagating the distribution of the high-dimensional state forward in time and from solving the high-dimensional inference problem when the new data is observed.

To reduce the computational complexity of filtering methods, our proposed dimension reduction method is applied in an \textit{offline-online} fashion. 
In the \textit{offline} phase, we identify a low-dimensional subspace of the state space before solving the inverse problem, using either the method of snapshots or regularized covariance estimation.  
In the \textit{online} phase, the computational complexity of various (Gaussian) filtering algorithms---including the Kalman filter, extended Kalman filter, and ensemble Kalman filter---is reduced by constraining the update and the prediction steps within the resulting subspace, in a unified subspace-constrained Bayesian framework.

%

The success of the proposed approach naturally requires that the unknown states can be captured by a low-dimensional basis. 
This is the case, for instance, if either the model states are sufficiently smooth or the states can be captured by a low-dimensional attractor. 
We show numerical examples where the reduction provides significant computational savings in different ways---by reducing the dimension of the linear systems involved in the prediction and update steps, but also, in ensemble filtering approaches, by reducing the number of ensemble members required. We discuss the limitations of the proposed approach in cases where the state cannot be represented efficiently in a fixed low-dimensional subspace. 
We also discuss several different ways to construct the subspace based on existing ``snapshots'' of the model states. 


The idea of reducing the dimension of filtering algorithms has been investigated before. 
The closest existing algorithm to our approach is the reduced-order Kalman filter (ROKF) \cite{cane96,hoteit04}, where dimension reduction is obtained by projecting the model dynamics onto a fixed low-dimensional subspace. 
While the ROKF shares some similarities with the present approach, fundamental differences remain: our algorithms do not explicitly project the model dynamics onto the subspace, \todo{check this} but rather constrain the inference (update) step using the subspace.  We show that the latter strategy yields more appropriate prior distributions for each inference problem.
Moreover, we extend the discussion of dimension reduction to ensemble filtering methods. Differences between the present approach and the ROKF are analyzed in detail later in the paper.

Another related approach is the reduced-rank Kalman filter (RRKF) of \cite{fisher98}. 
In this approach, one propagates prediction uncertainties only along directions in the state space where the variance of the states grows most quickly (the so-called Hessian singular vectors). 
%
%
The difference with our approach is that the subspace of RRKF is re-computed at each filtering step through the solution of an eigenvalue problem, whereas in our approach the basis is fixed and computed offline.  Thus, RRKF is a local-in-time approach, based on local linearization, whereas our approach tries to find a low-dimensional subspace for the state based on an analysis of global-in-time dynamics.
Our approach is less computationally intensive, but its applicability may be restricted to cases where a global low-dimensional representation for model states exists. Our numerical examples, however,  demonstrate that this simple strategy can yield significant computational savings in a range of filtering algorithms.

In \cite{chorin04}, dimension reduction is sought not in the Gaussian filtering context, but for a sequential Monte Carlo (particle filtering) method. Again, dimension reduction is performed locally in time. Filtering is restricted to coordinates spanning the most unstable modes around the current nominal trajectory; interestingly, for spatially distributed systems, these unstable modes are often the low-wavenumber components of the state.

Dimension reduction approaches for filtering problems can be beneficial in many ways. For instance, they reduce memory requirements, which are prohibitively large for high-dimensional problems when standard Kalman filters are applied. Indeed, memory constraints have motivated the development of various approximate filtering methods \cite{vkf,bardsley1}. The dimension reduction approach presented here reduces memory requirements, but also offers speedups that may be beneficial for (even smaller scale) real-time estimation and control problems, e.g., chemical process tomography \cite{aku09}. As noted above, speedups also extend to ensemble filtering methods: constraining the inference step onto a subspace implicitly regularizes the problem, and thus reduces the number of ensemble members required to achieve a given accuracy.

The paper is organized as follows. In Section \ref{sec:redu}, we review prior-based dimension reduction for static inverse problems and develop the linear-Gaussian filtering equations for subspace coordinates. In Section \ref{sec:nonlin}, we discuss how the method applies to extended Kalman filtering and ensemble methods. Section \ref{sec:subspace} discusses techniques for constructing the low-dimensional subspace. Section \ref{sec:rokf} analyzes the differences between our approach and the ROKF. In Sections~\ref{sec:numex}--\ref{sec:numex2}, we study the behavior of the dimension reduction approach via linear and nonlinear examples. Section \ref{sec:conc} offers some concluding remarks.

\section{Prior-based dimension reduction}
\label{sec:redu}

\subsection{Static problems}
\label{sec:static}

Our starting point is the prior-based dimension reduction technique for static inverse problems, which we briefly review here. 
The unknown function $x(s)$, $s \in \Omega$, is defined in some spatial domain $\Omega$. 
Discretizing $x(s)$ on a grid defined by a set of nodes $\{s_i\}_{i = 1}^{d}$ and some basis functions yields a $d$-dimensional vector $\mathbf{x} = [x(s_1), \ldots, x(s_d)]^\top \in \mathbb{R}^d$. 
The discretized unknown vector $\mathbf{x}$ is related to observations $\mathbf{y} \in \mathbb{R}^m$ via the model 
\begin{equation}
\mathbf{y}=f(\mathbf{x})+\varepsilon,
\end{equation}
where $\epsilon \sim \mathrm{N}(\mathbf{0},\mathbf{R})$ and $f$ is a (possibly nonlinear) mapping from the unknown $\mathbf{x}$ to the observable output.  Moreover, let us assume that we have a Gaussian prior $\mathbf{x} \sim \mathrm{N}(\bm{\mu},\mathbf{\Sigma})$. Then, the posterior density for $\mathbf{x}$ is
\begin{equation}
p(\mathbf{x} | \mathbf{y}) \propto \exp \left( -\frac{1}{2} \left( \Vert\mathbf{y}-f(\mathbf{x})\Vert_{\mathbf{R}}^2 + \Vert \mathbf{x}-\bm{\mu} \Vert_{\mathbf{\Sigma}}^2 \right) \right),
\end{equation}
where $\Vert\mathbf{b}\Vert_{\mathbf{A}}^2$ denotes the quadratic form $\mathbf{b}^\top\mathbf{A}^{-1}\mathbf{b}$.

The idea in prior-based dimension reduction is to constrain the problem onto a subspace that contains most of the variability allowed by the prior; see, for instance \cite{marzouk09}. This can be done by computing the singular value decomposition (SVD) of the prior covariance matrix'' $\mathbf{\Sigma} = \mathbf{U}\mathbf{\Lambda}\mathbf{U}^T$, where $\mathbf{U} \in \mathbb{R}^{d \times d}$ contains the singular vectors $\mathbf{u}_1,\mathbf{u}_2,\ldots,\mathbf{u}_d$ (as colums) and $\mathbf{\Lambda}=\mathrm{diag}(\lambda_1,\ldots,\lambda_d)$ has the singular vectors in the diagonal. Dimension is reduced by representing the unknown as a linear combination of the $r$ leading (scaled) singular vectors:
\begin{equation}
\mathbf{x}=\bm{\mu}+\mathbf{P}_r\bm{\alpha}\quad \mathrm{with}\ \mathbf{P}_r=\mathbf{U}_r\mathbf{\Lambda}_r^{1/2}=[\sqrt{\lambda}_1\mathbf{u}_1, \sqrt{\lambda}_2\mathbf{u}_2, \ldots, \sqrt{\lambda}_r\mathbf{u}_r].
\end{equation} 
Inserting this parameterization into the problem leads to the following posterior density for subspace coordinates $\bm{\alpha}$:
\begin{equation}
p(\bm{\alpha} | \mathbf{y}) \propto \exp \left( -\frac{1}{2} \left( \Vert\mathbf{y}-f(\bm{\mu}+\mathbf{P}_r\bm{\alpha})\Vert_{\mathbf{R}}^2 + \Vert \mathbf{P}_r\bm{\alpha} \Vert_{\mathbf{\Sigma}}^2 \right) \right).
\end{equation}
It is easy to verify that the prior term simplifies to $\Vert \mathbf{P}_r\bm{\alpha} \Vert_{\mathbf{\Sigma}}^2=\bm{\alpha}^\top (\mathbf{P}_r^\top \mathbf{\Sigma}^{-1} \mathbf{P}_r) \bm{\alpha} = \Vert\bm{\alpha}\Vert_{\mathbf{I}_r}^2$, where $\mathbf{I}_r$ is the $r \times r$ identity matrix. The posterior can be thus written simply as 
\begin{equation}
p(\bm{\alpha} | \mathbf{y}) \propto \exp \left( -\frac{1}{2} \left( \Vert\mathbf{y}-f(\bm{\mu}+\mathbf{P}_r\bm{\alpha})\Vert_{\mathbf{R}}^2 + \Vert \bm{\alpha} \Vert_{\mathbf{I}_r}^2 \right) \right),
\end{equation}
and the dimension of the inverse problem has been reduced from $d$ to $r$. This can be helpful, for instance, when MCMC samplers, which are challenging to apply in high-dimensional problems, are used to quantify uncertainty in the parameters. 

If the model is linear, $f(\mathbf{x})=\mathbf{Fx}$, the posterior is Gaussian $\mathrm{N}(\bm{\alpha}_{\mathrm{pos}},\mathbf{\Psi}_{\mathrm{pos}})$ with mean and covariance matrix given by 
\begin{eqnarray}
\mathbf{\Psi}_{\mathrm{pos}} & = & \left( (\mathbf{FP}_r)^\top\mathbf{R}^{-1}(\mathbf{FP}_r) + \mathbf{I}_r \right)^{-1} \\
\bm{\alpha}_{\mathrm{pos}} & = & \mathbf{\Psi}_{\mathrm{pos}}  (\mathbf{FP}_r)^\top\mathbf{R}^{-1} (\mathbf{y}-\mathbf{F}\bm{\mu}).
\end{eqnarray}
Thus, one needs to apply the model to the $r$ columns of $\mathbf{P}_r$ and solve $r$-dimensional linear system, which is computationally much easier than solving the full problem if $r \ll d$. 

\subsection{Dynamical problems}
\label{sec:dyn}

Here, we discuss how dimension reduction can be implemented for dynamical state estimation problems. Let us begin with the following linear Gaussian state space model:
\begin{eqnarray}
\label{ssmodel1}
\mathbf{x}_k & = & \mathbf{M}_k\mathbf{x}_{k-1} + \mathbf{E}_k \\
\mathbf{y}_k & = & \mathbf{H}_k\mathbf{x}_k + \mathbf{e}_k.
\label{ssmodel2}
\end{eqnarray}
In the above system, $\mathbf{M}_k \in \mathbb{R}^{d\times d}$ is the forward model that evolves the state in time and $\mathbf{H}_k \in \mathbb{R}^{m\times d}$ is the observation model that maps the state to the observations. The model and observation errors are assumed to be zero mean Gaussians: $\mathbf{E}_k \sim \mathrm{N}(\mathbf{0},\mathbf{Q}_k)$ and $\mathbf{e}_k \sim \mathrm{N}(\mathbf{0},\mathbf{R}_k)$ with known covariance matrices $\mathbf{Q}_k \in \mathbb{R}^{d\times d}$ and $\mathbf{R}_k \in \mathbb{R}^{m\times m}$.

The linear Gaussian problem can be solved with the Kalman filter, which proceeds sequentially as follows. Assume that, at time step $k$, the marginal posterior is the following Gaussian: 
\begin{equation}
\mathbf{x}_{k-1} | \mathbf{y}_{1:k-1} \sim \mathrm{N}(\mathbf{x}_{k-1}^a,\mathbf{C}_{k-1}^a).
\end{equation}
The {\it prediction step} involves propagating this Gaussian forward with the model $\mathbf{M}_k$, which yields the Gaussian
\begin{equation}
\mathbf{x}_k | \mathbf{y}_{1:k-1} \sim \mathrm{N}(\mathbf{x}_{k}^f,\mathbf{C}_{k}^f),
\end{equation}
where $\mathbf{x}_k^f = \mathbf{M}_k\mathbf{x}_{k-1}^a$ and $\mathbf{C}_k^f = \mathbf{M}_k\mathbf{C}_{k-1}^{a}\mathbf{M}_k^T+\mathbf{Q}_k$. { Throughout the paper, we follow the notation commonly used in weather forecasting and data assimilation literature: we use the superscript $a$ to refer to the ``analysis'' (posterior) estimate updated with the most recent observations, and the superscript $f$ to refer to the ``forecast'' (prior) estimate.}

In the {\it update step}, the predicted Gaussian is updated with the new observations that become available. The resulting posterior density is 
\begin{equation}
p(\mathbf{x}_k | \mathbf{y}_{1:k}) \propto \exp \left( -\frac{1}{2} \left( \Vert\mathbf{x}_k^f-\mathbf{x}_k\Vert_{\mathbf{C}_k^f}^2 + \Vert\mathbf{y}_k-\mathbf{H}_k\mathbf{x}_k\Vert_{\mathbf{R}_k}^2 \right) \right),
\label{fullpost}
\end{equation}
which is, again, Gaussian with known mean and covariance matrix given by the standard Kalman filter formulas, which we choose not to rewrite here {(see any standard textbook on the subject, e.g., \cite{sarkka13}).}

The most straightforward application of the prior-based dimension reduction technique in dynamical problems would be to define $\mathbf{P}_r$ separately for each filter step via the leading singular values and vectors of the prior covariance matrix $\mathbf{C}_k^f$. This approach has a few potential problems. First, computing the local leading singular vectors at each time step can be a computationally challenging task. Second, by truncating the prior covariance, we might discard directions that seem less important (i.e., that have low prior variance) locally in time, but that become relevant at a later time step. In our experiments, this approach led to inconsistent behavior of the filter; good filtering accuracy was obtained for some cases, but in other cases the filter performed poorly or even diverged.

Here, we examine an alternative, simpler strategy, where a global subspace is constructed \textit{a priori} (before the filtering is started) and is then fixed for the filtering. This approach is motivated by the fact that the state of a dynamical system often lives in a subspace of much smaller dimension than the full state space; the state vector often has some properties (e.g., smoothness) that enable it to be effectively described in a low-dimensional subspace. If we can capture the subspace where the essential dynamics of the system happen, we can potentially reduce the whole filtering procedure onto this subspace. This idea is discussed here. 

Now, we parameterize the unknown as  $\mathbf{x}_{k}=\mathbf{x}_{k}^f+\mathbf{P}_r\bm{\alpha}_{k}$, where $\mathbf{P}_r \in \mathbb{R}^{d\times r}$ is a fixed reduction operator that does not change in time and $\mathbf{x}_{k}^f$ is the predicted (prior) mean. For now, we assume that such a representation exists; discussion about how to construct $\mathbf{P}_r$ is reserved for Section \ref{sec:subspace}. To derive the filtering equations for the subspace coordinates $\bm{\alpha}_k$, assume that the marginal posterior distribution for $\bm{\alpha}_{k-1}$ at time $k-1$ is
\begin{equation}
\bm{\alpha}_{k-1} | \mathbf{y}_{1:k-1} \sim \mathrm{N}(\bm{\alpha}_{k-1}^a, \mathbf{\Psi}_{k-1}^a).
\end{equation} 
Transforming this Gaussian distribution to the original coordinates using $\mathbf{x}_{k-1}=\mathbf{x}_{k-1}^f+\mathbf{P}_r\bm{\alpha}_{k-1}$ yields the following Gaussian distribution in the full state space:
\begin{equation}
\mathbf{x}_{k-1} | \mathbf{y}_{1:k-1} \sim \mathrm{N}(\mathbf{x}_{k-1}^f+\mathbf{P}_r\bm{\alpha}_{k-1}^a,\mathbf{P}_r\mathbf{\Psi}_{k-1}^a\mathbf{P}_r^\top).
\end{equation}
By propagating this Gaussian distribution forward with $\mathbf{M}_k$, we obtain the mean and the covariance matrix of the predictive distribution $\mathbf{x}_k | \mathbf{y}_{1:k-1} \sim \mathrm{N}(\mathbf{x}_k^f,\mathbf{C}_k^f)$ in the original coordinates: 
\begin{eqnarray}
\mathbf{x}_k^f & = & \mathbf{M}_k(\mathbf{x}_{k-1}^f+\mathbf{P}_r\bm{\alpha}_{k-1}^a)\\
\mathbf{C}_k^f & = & (\mathbf{M}_k\mathbf{P}_r)\mathbf{\Psi}_{k-1}^{a}(\mathbf{M}_k\mathbf{P}_r)^T+\mathbf{Q}_k.
\end{eqnarray}
Applying this as the prior, and inserting the parameterization $\mathbf{x}_{k}=\mathbf{x}_k^f+\mathbf{P}_r\bm{\alpha}_{k}$ into (\ref{fullpost}) yields the following marginal posterior density for $\bm{\alpha}_k$:
\begin{equation}
p(\bm{\alpha}_k | \mathbf{y}_{1:k})   
 \propto \exp \left( -\frac{1}{2} \left( \Vert\mathbf{P}_r\bm{\alpha}_k\Vert_{\mathbf{C}_k^f}^2 + \Vert\mathbf{y}_k-\mathbf{H}_k\mathbf{x}_k^f-\mathbf{H}_k\mathbf{P}_r\bm{\alpha}_k\Vert_{\mathbf{R}_k}^2 \right) \right).
\label{post}
\end{equation}
{ This is equivalent to a linear problem with Gaussian likelihood $\mathbf{y}_k-\mathbf{H}_k\mathbf{x}_k^f \sim \mathrm{N}(\mathbf{H}_k\mathbf{P}_r\bm{\alpha}_k,\mathbf{R}_k)$ and zero mean Gaussian prior $\bm{\alpha}_k \sim \mathrm{N}(\mathbf{0},(\mathbf{P}_r^\top(\mathbf{C}_k^f)^{-1}\mathbf{P}_r)^{-1})$.} The resulting posterior is thus $\bm{\alpha}_k | \mathbf{y}_{1:k} \sim \mathrm{N}(\bm{\alpha}_k^a,\mathbf{\Psi}_k^a)$ where
\begin{eqnarray}
\label{est1}
\mathbf{\Psi}_k^a & = & \left( (\mathbf{H}_k\mathbf{P}_r)^\top\mathbf{R}_k^{-1}(\mathbf{H}_k\mathbf{P}_r)+\mathbf{P}_r^\top(\mathbf{C}_k^f)^{-1}\mathbf{P}_r \right)^{-1}, \\
\bm{\alpha}_k^a & = & \mathbf{\Psi}_k^a (\mathbf{H}_k\mathbf{P}_r)^\top\mathbf{R}_k^{-1}\mathbf{r}_k,
\label{est2}
\end{eqnarray}
and $\mathbf{r}_k=\mathbf{y}_k-\mathbf{H}_k\mathbf{x}_k^f$ is the prediction residual. Note that now $\mathbf{P}_r$ does not whiten the prior, in contrast with the prior-based dimension reduction discussed in Section \ref{sec:static} for static problems, and thus the term $\mathbf{P}_r^\top(\mathbf{C}_k^f)^{-1}\mathbf{P}_r$ is not equal to the identity. Moreover, the matrix $\mathbf{C}_k^f$ cannot be formed explicitly in high-dimensional problems. To efficiently evaluate $(\mathbf{C}_k^f)^{-1}\mathbf{P}_r$, we recall that here
\begin{equation}
\mathbf{C}_{k}^f=(\mathbf{M}_k\mathbf{P}_r)\mathbf{\Psi}_{k-1}^{a}(\mathbf{M}_k\mathbf{P}_r)^T+\mathbf{Q}_k=(\mathbf{M}_k\mathbf{P}_r\mathbf{A}_k)(\mathbf{M}_k\mathbf{P}_r\mathbf{A}_k)^T+\mathbf{Q}_k,
\end{equation}
where $\mathbf{A}_k \in \mathbb{R}^{r \times r}$ is the matrix square root $\mathbf{\Psi}_{k-1}^{a}=\mathbf{A}_k\mathbf{A}_k^T$. To shorten the notation, let us denote $\mathbf{B}_k=\mathbf{M}_k\mathbf{P}_r\mathbf{A}_k \in \mathbb{R}^{d \times r}$. Now, applying the Sherman-Morrison-Woodbury matrix inversion formula yields 
\begin{equation}
(\mathbf{C}_{k}^f)^{-1}=(\mathbf{B}_k\mathbf{B}_k^T+\mathbf{Q}_k)^{-1} = \mathbf{Q}_k^{-1}-\mathbf{Q}_k^{-1}\mathbf{B}_k(\mathbf{B}_k^T\mathbf{Q}_k^{-1}\mathbf{B}_k+\mathbf{I}_r)^{-1}\mathbf{B}_k^T\mathbf{Q}_k^{-1},
\label{smw}
\end{equation}
where $\mathbf{I}_r$ is the $r \times r$ identity matrix. Now, the matrix $\mathbf{B}_k^T\mathbf{Q}_k^{-1}\mathbf{B}_k+\mathbf{I}_r$ that needs to be inverted is only $r \times r$. Thus, a product $(\mathbf{C}_k^f)^{-1}\mathbf{b}$ can be efficiently computed as long as $\mathbf{Q}_k^{-1}\mathbf{b}$ is easy to compute. This condition must hold in order for this technique to work. In practice, $\mathbf{Q}_k$ is usually a simple (e.g., diagonal) matrix postulated by the user. 

To sum up, a single step of the reduced Kalman filter with a fixed basis is given as an algorithm below. Assume that we have the previous estimate $\bm{\alpha}_{k-1}^{a}$ and its covariance matrix $\mathbf{\Psi}_{k-1}^{a}$ available. Then the algorithm reads as follows:\\
\ \\
\fbox{\parbox{0.9\linewidth}{
\textbf{Algorithm 1: one step of the reduced Kalman filter.}\\ {\it Input:} $\bm{\alpha}_{k-1}^{a}$ and $\mathbf{\Psi}_{k-1}^{a}$. {\it Output:} $\bm{\alpha}_k^a$ and $\mathbf{\Psi}_{k}^{a}$. 
\begin{enumerate}
\item Compute the prior mean $\mathbf{x}_k^f=\mathbf{M}_k(\mathbf{x}_{k-1}^f+\mathbf{P}_r\bm{\alpha}_{k-1}^{a})$.
\item Perform the decomposition $\mathbf{\Psi}_{k-1}^{a}=\mathbf{A}_k\mathbf{A}_k^\top$.
\item Compute the matrix $\mathbf{B}_k=\mathbf{M}_k\mathbf{P}_r\mathbf{A}_k$.
\item Compute $(\mathbf{C}_k^f)^{-1}\mathbf{P}_r$ via the matrix inversion formula (\ref{smw}).
\item Compute $\bm{\alpha}_k^a$ and $\mathbf{\Psi}_{k}^{a}$ via formulas (\ref{est1})--(\ref{est2}).\\
\end{enumerate}
}}
\ \\\\

\begin{remark}
Computationally, this version is much lighter than the standard Kalman filter if $r \ll d$. In the standard Kalman filter, the prediction covariance matrix is computed as $\mathbf{C}_{k}^f = \mathbf{M}_k\mathbf{C}_{k-1}^{a}\mathbf{M}_k^T+\mathbf{Q}_k$; that is, we need to compute products of $d \times d$ matrices (or to apply the forward model to the $d$ columns of $\mathbf{C}_{k-1}^a$). Moreover, when updating the prior covariance, one needs to operate with $\mathbf{C}_{k}^f$ and solve a system of $m$ linear equations. In the approach described here, one needs to work with $d \times r$ matrices $\mathbf{B}_k$, solve a system of $r$ linear equations, and do one inversion of a $r \times r$ matrix. Also, here the basis vectors are fixed, so we avoid solving local eigenvalue problems, which are needed, for instance, in the reduced rank Kalman filter of \cite{fisher98}.
\end{remark}

\begin{remark}
Here, the parameterization is centered at the predicted mean, $\mathbf{x}_k=\mathbf{x}_k^f+\mathbf{P}_r\bm{\alpha}_k$. An alternative would be to use a fixed mean, $\mathbf{x}_k=\bm{\mu}+\mathbf{P}_r\bm{\alpha}_k$, where $\bm{\mu}$ is some fixed offset. The former looks for a correction to the predicted mean in the subspace, whereas the latter attempts to describe the state vector itself in a fixed subspace. In our experiments, the former yields much better filter accuracy, especially with small $r$. 
\end{remark}

\section{Extensions to nonlinear problems}
\label{sec:nonlin}

Here, we discuss two ways to extend the dimension reduction idea to problems where the evolution and/or observation models are nonlinear. We still assume additive Gaussian model and observation errors, so our state space model now reads as
\begin{eqnarray}
\label{ssmodel1_nl}
\mathbf{x}_k & = & \mathcal{M}(\mathbf{x}_{k-1}) + \mathbf{E}_k \\
\mathbf{y}_k & = & \mathcal{H}(\mathbf{x}_k) + \mathbf{e}_k,
\label{ssmodel2_nl}
\end{eqnarray}
where $\mathcal{M}$ and $\mathcal{H}$ are the nonlinear forward and observation models. We start by discussing the extended Kalman filter, which requires linearizations of the forward and observation models. Then, we discuss ensemble filtering techniques where linearizations are not needed. 

\subsection{Extended Kalman filtering}
\label{sec:ekf}

The extended Kalman filter (EKF) replaces the model and observation matrices in the KF with their linearized versions. Thus, the algorithm is the same as Algorithm~1 in Section \ref{sec:redu}, but the mean is propagated with the nonlinear forward model and the prediction residual is calculated with the nonlinear observation model. 
Elsewhere $\mathbf{M}_k$ and $\mathbf{H}_k$ are replaced with 
\begin{equation}
\mathbf{M}_k = \left. \frac{\partial \mathcal{M}(\mathbf{x}_{k-1})}{\partial \mathbf{x}_{k-1}} \right|_{\mathbf{x}_{k-1}=\mathbf{x}_{k-1}^{a}} \quad
\mathbf{H}_k = \left. \frac{\partial \mathcal{H}(\mathbf{x}_{k})}{\partial \mathbf{x}_{k}} \right|_{\mathbf{x}_{k}=\mathbf{x}_{k}^{f}}.
\end{equation}
Note that for large scale problems, computing the above matrices explicitly is not feasible. Instead, one often derives the linearized model analytically (as an operator), which makes it possible to propagate vectors forward with the linear model, which is equivalent to computing products $\mathbf{M}_k \mathbf{b}$ where $\mathbf{b} \in \mathbb{R}^{d \times 1}$. Here, these \textit{tangent linear codes} need to be applied in steps~(iii)~and~(v) of Algorithm~1. In step~(iii), we compute $\mathbf{B}_k=\mathbf{M}_k\mathbf{P}_r\mathbf{A}_k$, which can be done by applying the linearized forward model to the $r$ columns of $\mathbf{P}_r\mathbf{A}_k$. In step~(v), we need $\mathbf{H}_k\mathbf{P}_r$, which can be computed by applying the linearized observation model to the $r$ columns of $\mathbf{P}_r$. In both cases, we only need to propagate $r$ vectors through the linearized models, instead of $d$ vectors as in the standard EKF. 

\subsection{Ensemble filtering}
\label{sec:enkf}

Ensemble filters have become popular for solving very high-dimensional dynamical state estimation problems arising in geophysical applications such as numerical weather prediction. The development started started from the ensemble Kalman filter (EnKF, \cite{enkf,enkf2}) in the 1990s, and different variants are under active development. The idea of the EnKF is to represent the state and its uncertainty with samples (an ``ensemble'' of states), and, roughly speaking, to replace the covariances in the filtering formulas with their empirical estimates calculated from the samples.

{ For high-dimensional problems that involve complex physical models, the ensemble size is necessarily much smaller than the dimension of the problem. As a result, the obtained covariance estimates are rank-deficient and can suffer from ``spurious correlations'' (unphysical correlations appearing randomly due to small sample size); see, e.g., \cite{eakf,evensen07,hamill01} for discussion. To overcome these issues, various \textit{localization} techniques have been proposed, where the empirical covariance estimates are regularised by, for instance, explicitly removing unrealistic distant correlations from the covariance matrices; see \cite{anderson03,greybush11,ott04}. Recently, adaptive localization techniques have also been developed, where the localization mechanism is tuned on-line in the filter \cite{bishop11,kirchgessner14}. Localization is one of the key techniques to make EnKFs work for small ensemble sizes.

In addition to rank deficiency, the classical EnKF suffers from sampling error, as the observation and model errors are accounted for by randomly perturbing the observations and predicted ensemble members during the estimation. To avoid this additional variance in the resulting estimates, so-called square root EnKFs have been developed, which are deterministic schemes where no such random perturbations are used for the observations \cite{eakf,etkf,srf}. Accounting for model error still remains a difficulty, although some techniques have been recently proposed; see, e.g., \cite{raanes15} and the references therein. 

In the method discussed here, the state is constrained onto a subspace, which heavily regularises the estimation problem; for instance, if the state is smooth, the rough features are explicitly removed from the estimation problem, and all of the information in the samples can be used for inferring the smooth features. As a result, the need for localization is diminished, as demonstrated in the numerical examples in Section \ref{sec:numex}. Moreover, model error can be can be easily included in the approach, provided that we are able to apply the model error covariance matrix to a vector efficiently.} 



Here, we present one way of extending the dimension reduction idea to ensemble filtering. Let us now consider the state space model where the forward model is nonlinear but the observation model is linear (the nonlinear observation model case is discussed later):
\begin{eqnarray}
\label{ssmodel1_linobs}
\mathbf{x}_k & = & \mathcal{M}(\mathbf{x}_{k-1}) + \mathbf{E}_k \\
\mathbf{y}_k & = & \mathbf{H}_k\mathbf{x}_k + \mathbf{e}_k.
\label{ssmodel2_linobs}
\end{eqnarray}
In ensemble filtering, we represent the distribution of $\bm{\alpha}_k$ with samples. Let us assume that at time step $k-1$ we have $N_{\mathrm{ens}}$ posterior samples $\{\bm{\alpha}_{k-1}^{a,1},\bm{\alpha}_{k-1}^{a,2},\ldots,\bm{\alpha}_{k-1}^{a,N_{\mathrm{ens}}} \}$ available, sampled from the Gaussian posterior $\mathrm{N}(\bm{\alpha}_{k-1}^{a},\mathbf{\Psi}_{k-1}^a)$. These obviously correspond to samples $\mathbf{x}_{k-1}^{a,i}=\mathbf{x}_{k-1}^f+\mathbf{P}_r\bm{\alpha}_{k-1}^{a,i}$ in the full state space. In the prediction step, we move the posterior samples forward with the dynamical model: $\mathbf{x}_{k}^{f,i}=\mathcal{M}(\mathbf{x}_{k-1}^{a,i})$. Then, we compute the empirical covariance matrix of the predicted samples and add the model error covariance to obtain the prediction error covariance:
\begin{equation}
\mathbf{C}_k^f=\mathrm{Cov} \left(  \mathcal{M}(\mathbf{x}_{k-1}) + \mathbf{E}_k \right) \approx \mathbf{X}_k\mathbf{X}_k^\top + \mathbf{Q}_k,
\end{equation}
where 
\begin{equation}
\mathbf{X}_k=\left[ (\mathbf{x}_{k}^{f,1}-\mathbf{x}_{k}^{f}) \quad (\mathbf{x}_{k}^{f,2}-\mathbf{x}_{k}^{f}) \ \ldots \ (\mathbf{x}_{k}^{f,N_{\mathrm{ens}}}-\mathbf{x}_{k}^{f}) \right] / \sqrt{N_{\mathrm{ens}}} \in \mathbb{R}^{d \times N_{\mathrm{ens}}}.
\label{Xk}
\end{equation}
Thus, $\mathbf{X}_k\mathbf{X}_k^\top$ is the empirical covariance estimate computed from the prediction ensemble. The mean $\mathbf{x}_{k}^{f}$ is taken to be the posterior mean from the previous step propagated via the evolution model, $\mathbf{x}_{k}^{f}=\mathcal{M}(\mathbf{x}_{k-1}^{a})=\mathcal{M}(\mathbf{x}_{k-1}^f+\mathbf{P}_r\bm{\alpha}_{k-1}^{a})$, instead of the empirical mean computed from the prediction ensemble, which is why we divide by $\sqrt{N_{\mathrm{ens}}}$ instead of $\sqrt{N_{\mathrm{ens}}-1}$, see the remarks below for more discussion about this choice. 

If the observation model is linear, the reduced dimension ensemble filtering algorithm stays almost the same as the reduced KF algorithm (Algorithm~1 in Section~\ref{sec:redu}). The only difference is the prior covariance matrix $\mathbf{C}_k^f$, which is now defined via the prediction ensemble. When $N_{\mathrm{ens}} \ll d$, operating with $(\mathbf{C}_k^f)^{-1}$, needed in step~(v) of Algorithm~1, can still be done efficiently via the Sherman-Morrison-Woodbury inversion formula:
\begin{equation}
(\mathbf{C}_k^f)^{-1}=(\mathbf{X}_k\mathbf{X}_k^\top + \mathbf{Q}_k)^{-1}=\mathbf{Q}_k^{-1}-\mathbf{Q}_k^{-1}\mathbf{X}_k(\mathbf{X}_k^T\mathbf{Q}_k^{-1}\mathbf{X}_k+\mathbf{I}_{N_{\mathrm{ens}}})^{-1}\mathbf{X}_k^T\mathbf{Q}_k^{-1}.
\label{smw2}
\end{equation}
Note that now, when applying $(\mathbf{C}_k^f)^{-1}$ to a vector, we are left with the inversion of an $N_{\mathrm{ens}} \times N_{\mathrm{ens}}$ matrix instead of an $r \times r$ matrix. 

To summarise, one step of the ensemble Kalman filter with dimension reduction is given below.\\
\ \\
\fbox{\parbox{0.9\linewidth}{
\textbf{Algorithm 2: one step of the ensemble Kalman filter with reduced dimension.} {\it Input:} $\bm{\alpha}_{k-1}^{a}$ and $\mathbf{\Psi}_{k-1}^{a}$. {\it Output:} $\bm{\alpha}_k^a$ and $\mathbf{\Psi}_{k}^{a}$. 
\begin{enumerate}
\item Draw $N_{\mathrm{ens}}$ samples $\{\bm{\alpha}_{k-1}^{a,i}\}$ from $\mathrm{N}(\bm{\alpha}_{k-1}^{a},\mathbf{\Psi}_{k-1}^{a})$ and transform samples into the full state space: $\mathbf{x}_{k-1}^{a,i}=\mathbf{x}_{k-1}^f+\mathbf{P}_r\bm{\alpha}_{k-1}^{a,i}$.
\item Compute the prior mean $\mathbf{x}_k^f=\mathcal{M}(\mathbf{x}_{k-1}^f+\mathbf{P}_r\bm{\alpha}_{k-1}^{a})$ and propagate the samples: $\mathbf{x}_{k}^{f,i}=\mathcal{M}(\mathbf{x}_{k-1}^{a,i})$.
\item Form $\mathbf{X}_k=\left[ (\mathbf{x}_{k}^{f,1}-\mathbf{x}_{k}^{f}) \quad (\mathbf{x}_{k}^{f,2}-\mathbf{x}_{k}^{f}) \ \ldots \ (\mathbf{x}_{k}^{f,N_{\mathrm{ens}}}-\mathbf{x}_{k}^{f}) \right] / \sqrt{N_{\mathrm{ens}}}$.
\item Compute $(\mathbf{C}_k^f)^{-1}\mathbf{P}_r$ via the matrix inversion formula (\ref{smw2}).
\item Compute $\bm{\alpha}_k^a$ and $\mathbf{\Psi}_{k}^{a}$ via formulas (\ref{est1})--(\ref{est2}).\\
\end{enumerate}
}}
\ \\\\

\begin{remark}
The computational cost of the ensemble algorithms is dictated by both the number of basis vectors $r$ and the number of ensemble members $N_{\mathrm{ens}}$. The computational cost is similar to that of the standard EnKF. What makes dimension reduction attractive from ensemble filtering point of view is that the number of samples needed to capture the distribution of the $r$-dimensional variable $\bm{\alpha}_k$ can be much smaller than the number of samples needed to get accurate filtering results for the $d$-dimensional variable $\mathbf{x}_k$ in the full space. Thus, similar performance can be obtained with fewer ensemble members, as can be observed in the numerical examples in Sections \ref{sec:numex}--\ref{sec:numex2}. 
\end{remark}

\begin{remark}
The ensemble filter presented above differs from the classical ensemble Kalman filter (EnKF) developed in \cite{enkf,enkf2}. The classical EnKF is a non-Gaussian filter; it applies a linear update with perturbed observations to non-Gaussian prediction samples to get the posterior ensemble. The version presented here is a Gaussian filter in the sense that the prior is assumed to be a Gaussian whose covariance matrix is estimated from the prediction ensemble. 
\end{remark}

\begin{remark}
{ Another difference between the proposed method and many other ensemble filters is that the prior mean here is $\mathbf{x}_{k}^{f}=\mathcal{M}(\mathbf{x}_{k-1}^{a})$ instead of the empirical ensemble mean used, for instance, in the classical EnKF. Technically, one could easily choose the ensemble mean as $\mathbf{x}_{k}^{f}$ as well. However, we have noticed that the choice $\mathbf{x}_{k}^{f}=\mathcal{M}(\mathbf{x}_{k-1}^{a})$ for propagating the mean, analogous to EKF and variational (3D-Var and 4D-Var) methods, works better for many problems. One reason might be that the $\mathbf{x}_{k}^{f}$ obtained this way lies close to the attractor of the forward model, whereas the sample mean obtained from a small number of ensemble members might be further away from it and thus represent an ``unphysical'' state. A similar approach was taken in some recently developed filtering algorithms; see, for instance, \cite{solonen14} for some discussion.}
\end{remark}

\begin{remark}
If the observation model is nonlinear, the posterior distribution for $\bm{\alpha}_k$ is no longer Gaussian. Then, one way forward is to apply the linearized observation operator as in the EKF to obtain a Gaussian approximation of the posterior, and to sample new members from the Gaussian. Another way is to sample new posterior samples for $\bm{\alpha}_k$ directly from the non-Gaussian posterior
\begin{equation}
p(\bm{\alpha}_k | \mathbf{y}_{1:k}) \propto \exp \left( -\frac{1}{2} \left( \Vert\mathbf{P}_r\bm{\alpha}_k\Vert_{\mathbf{C}_k^f}^2 + \Vert\mathbf{y}_k-\mathcal{H}(\mathbf{x}_k^f+\mathbf{P}_r\bm{\alpha}_k)\Vert_{\mathbf{R}_k}^2 \right) \right)
\end{equation}
using, for instance, Markov chain Monte Carlo (MCMC) techniques, which should be feasible if $r$ is not too large and $\mathcal{H}$ is relatively simple (note that evaluating the posterior density does not require the forward model $\mathcal{M}$). For instance, novel optimization-based sampling techniques like \cite{bardsley14} (which requires Gaussian priors as above) \todo{I added this parenthetical statement as an alternative to citing the random-map implicit filter, which could be a distraction.}
can potentially be used to generate posterior samples efficiently, as discussed in \cite{solonen14} in connection with high-dimensional filtering problems.
\end{remark}

\begin{remark}
The subspace representation also opens up a way to implement other sample-based filtering techniques for high-dimensional problems, such as the popular unscented Kalman filter (see, e.g., \cite{sarkka13}), which, in the subspace version, would require $2r+1$ samples to propagate the covariance forward instead of $2d+1$ as in the full state space version. Even particle filtering in the subspace might be possible with a reasonable number of particles. These ideas are left for future study and not pursued further here. 
\end{remark}

\section{Reduced subspace construction}
\label{sec:subspace}

The reduced subspace basis $\mathbf{P}_r$ for representing the unknown state $\mathbf{x}_k$ is constructed in a similar way as for static problems discussed in Section \ref{sec:static}. 
Consider the covariance matrix $\mathbf{\Sigma}$ that represents the covariance structure of states, and in particular its eigendecomposition $\mathbf{\Sigma}=\mathbf{U \Lambda U}^\top$. We compute the basis $\mathbf{P}_r$ from the $r$ leading eigenvectors $\mathbf{U}_r=(\mathbf{u}_1,\ldots,\mathbf{u}_r)$ and square roots of the corresponding eigenvalues $\mathbf{\Lambda}_r=\mathrm{diag} (\lambda_1,\ldots,\lambda_r)$:
\begin{equation}
\mathbf{P}_r=\mathbf{U}_r\mathbf{\Lambda}_r^{1/2}.
\label{eq:repr}
\end{equation}
If the eigenvalues of the covariance matrix $\mathbf{\Sigma}$ decay quickly, the variation of states can be captured by a low-dimensional subspace spanned by the leading eigenvectors. 
We note that scaling the basis by the eigenvalues in (\ref{eq:repr}) is not necessary, but can unify the scales of the different state variables. 
Of course, this method requires access to $\mathbf{\Sigma}$.
In the rest of this section, we present several ways to estimate this covariance matrix.

\subsection{Principal component analysis}
\label{sec:pca}

Principal component analysis (PCA) can be applied to ``snapshots''---which are possible model states obtained from existing model simulations---for constructing low-dimensional subspaces of high-dimensional dynamical systems. 
Depending on the field of application, this procedure is also named empirical orthogonal functions \cite{preisendorfer88} in meteorology, or proper orthogonal decomposition \cite{sirovich87, everson95} in model reduction. 
The reduced basis obtained from PCA can then be used in either model reduction \cite{willcox02, rowley04} or filtering (e.g., the ROKF method \cite{cane96} or our approach as presented in Section~\ref{sec:dyn}).
%



In non-stationary inverse problems, our main interest is dynamical systems without steady states (e.g., chaotic models). 
In this setting, we use trajectories obtained from either a sufficiently long free model simulation or multiple model simulations with randomized initial conditions. 
Given a sufficient number of snapshots $\{ \mathbf{x}^{(i)} \}_{i=1}^{N}$, the subspace basis is computed using  (\ref{eq:repr}) via the eigendecomposition of empirical state covariance  \begin{equation}
\mathbf{\Sigma}=\frac{1}{N-1} \sum_{i=1}^N (\mathbf{x}^{(i)}-\overline{\mathbf{x}})(\mathbf{x}^{(i)}-\overline{\mathbf{x}})^\top,  \quad \overline{\mathbf{x}} = \frac{1}{N}  \sum_{i=1}^N \mathbf{x}^{(i)},
\end{equation}
where $\overline{\mathbf{x}}$ is the empirical state mean. 
For high-dimensional dynamical systems, it is not feasible to form the empirical covariance directly and apply dense matrix eigendecomposition methods.
In this case, either Krylov subspace methods \cite{golub12, lehoucq98} or randomized methods \cite{halko11, liberty07} should be applied together with matrix-free operations---the matrix vector product with $\mathbf{\Sigma}$---to compute the eigendecomposition. 

The number of snapshots naturally can affect the quality of the basis $\mathbf{P}_r$, and a sufficiently large $N$ should be chosen to capture the essential behavior of the model. 
Above, the basis is constructed from un-regularised empirical covariance estimates, without assuming any particular form for the covariance. 
%
If the number of snapshots available is limited, we can instead employ regularised covariance estimation techniques, where certain assumptions about the covariance structure are introduced to regularise the estimation problem. These techniques are discussed below.




\subsection{Regularized covariance estimation} 

Another viable route for state covariance estimation is to infer the state correlation structure from snapshots and {\it a priori} information such as smoothness assumptions. Here, we discuss how Gaussian processes (GPs) can be used for the task.
We consider two types of GPs: a stationary GP modeled by a kernel function \cite{rasmussen06}, or a (possibly) non-stationary GP modeled by a a differential operator \cite{rue05, lindgren11}.

\subsubsection{Stationary GPs via covariance kernels.}

We first discuss the kernel approach, where each element of the covariance matrix is given by a kernel function $k$ of the form
\begin{equation}
\mathbf{\Sigma}_{ij}(\bm{\theta})=k(s_i, s_j;\bm{\theta}),
\end{equation}
where $s_i, s_j \in \Omega$ are spatial locations used to discretize the states, and all the information about smoothness, correlation length, and variability can be encoded in the parameter $\bm{\theta}$. 
For example, one commonly used kernel function is the squared exponential kernel
\begin{equation}
k(s_i, s_j; \bm{\theta})=\theta_1 \exp \left( -\left(\frac{d(s_{i},s_{j})}{\theta_2} \right)^{2} \right),
\label{squared-exp}
\end{equation}
where $\theta_1$ and $\theta_2$ control the variability and correlation length, respectively, and $d(s_i,s_j)$ is a distance between points $s_i$ and $s_j$. With this kernel function, the eigenvalues decay quickly (exponentially), and it is easy to capture the kernel with a finite basis.

Given the mean of the GP, $\bm{\mu}$, empirically estimated from the snapshots, we estimate the parameters $\bm{\theta}$ in a Bayesian framework. 
The likelihood function of $\bm{\theta}$, given a snapshot collection $\{ \mathbf{x}^{(i)} \}_{i=1}^{N}$, takes the form
\begin{equation}
p\left(\mathbf{x}^{(1)},\ldots, \mathbf{x}^{(N)}  \vert  \bm{\theta} \right) = \frac{1}{\sqrt{(2\pi)^d |\mathbf{\Sigma}(\bm{\theta})|}} \exp \left( - \frac12 \sum_{i=1}^N (\mathbf{x}^{(i)}-\bm{\mu})^\top \mathbf{\Sigma}(\bm{\theta})^{-1} (\mathbf{x}^{(i)}-\bm{\mu}) \right).
\label{eq:like}
\end{equation} 
Combining the likelihood with a prior distribution $p(\bm{\theta})$, we obtain the {\it maximum a posteriori} estimate 
\begin{equation}
\hat{\bm{\theta}} = \mathrm{arg\,max}_{\bm{\theta}} \; p\left(\mathbf{x}^{(1)},\ldots, \mathbf{x}^{(N)} | \bm{\theta} \right) \times p(\bm{\theta}) 
\end{equation}
of the kernel parameters. 
The reduced subspace basis can then be computed from the eigendecomposition of the state covariance $\mathbf{\Sigma}(\hat{\bm{\theta}})$.

%

\begin{remark}
Here the smoothness assumption is used to ``fill the gap'' between the unknown high-dimensional correlations of the state and the information provided by a limited number of snapshots.
For a GP defined by stationary kernels, this assumption can be enforced by using a smooth kernel such as the squared exponential kernel. 
\end{remark}

\subsubsection{Nonstationary GPs via Gaussian Markov random fields.}

{
The stationary assumption we used in the above-mentioned kernel method may not be suitable for dynamical systems where the states have heterogeneous spatial correlations. 
Furthermore, operations with the dense covariance matrix $\mathbf{\Sigma}(\bm{\theta})$ in (\ref{eq:like}) can be computationally challenging for high-dimensional states, because the covariance matrix can be singular and computational costs of dense matrix operations---especially factorization and inversion---scale poorly with dimension. 

%

\newcommand{\precision}{\mathbf{\Omega}}
\newcommand{\mass}{\mathbf{\Delta}}
\newcommand{\stiff}{\mathbf{K}}

This motivates us to model the GP using the inverse of the covariance, i.e., the precision matrix $\precision = \mathbf{\Sigma}^{-1}$, so that Gaussian Markov random field (GMRF) models \cite{rue05} can be used to construct the precision as a sparse matrix. 
We particularly mention the work of \cite{lindgren11}, in which the sparse precision matrix is constructed from the finite element discretization of the following stochastic partial differential equation (SPDE):
\begin{equation} 
\label{eq:SPDE}
 \left(  \gamma(s) - \nabla \cdot (\kappa(s) \nabla)   \right)^{\frac{\alpha}{2}} x(s) = \mathcal{W}(s),
\end{equation}
where $\mathcal{W}(s)$ is a white noise process in space. 
For spatially constant $\gamma(s) = \gamma$ and $\kappa(s) = \kappa$, it can be shown that the solution $u(s)$ of the SPDE (\ref{eq:SPDE}) defines a Gaussian process with the Mat\'{e}rn family of correlation functions; see \cite{lindgren11} and references therein. 
The functions $\gamma(s)$ and $\kappa(s)$ together control the correlation length and variability of the GP, and the scalar $\alpha$ controls the smoothness, and therefore a nonstationary GP can be defined by prescribing spatially heterogeneous $\gamma(s)$ and $\kappa(s)$. 


For a positive integer $\alpha$, discretizing the SPDE (\ref{eq:SPDE}) yields a sparse precision matrix $\precision$. 
Here, we discretize (\ref{eq:SPDE}) using the finite element method with linear basis functions, 
\[
x(s) = \sum_{j = 1}^{J} x_j \phi_j(s),
\]
where $x_j = x(s_j),\ j = 1,\ldots, J$, is a set of nodal points and $\phi_j(s),\ j = 1,\ldots, J$, is a set of linear basis functions associated with the nodal points. 
Here the nodes are defined to coincide with the locations of the states in the filtering problem. 
The precision matrix $\precision$ can be constructed given the mass matrix and the stiffness matrix of the finite element discretization, which are defined as
\[
\mass_{ij} = \int \gamma(s) \phi_i(s) \phi_j(s) ds, \quad {\rm and} \quad \stiff_{ij} = \int \left[ \nabla \cdot (\kappa(s) \nabla) \phi_i(s) \right] \phi_j(s) ds, 
\]
where $i,j = 1,\ldots,J$.
The functions $\gamma(s)$ and $\kappa(s)$ can also be discretized by linear basis functions, which are given as
\[
\gamma(s) = \sum_{j = 1}^{J} \gamma_j \phi_j(s), \quad {\rm and} \quad \kappa(s) = \sum_{j = 1}^{J} \kappa_j \phi_j(s),  
\] 
respectively. 
This yields the local mass and stiffness matrices
\begin{eqnarray*}
\mass^{k}_{ij} & = & \int \phi_k(s) \phi_i(s) \phi_j(s) ds, \\
\stiff^{k}_{ij} & = & \int \left[ \nabla \cdot (\phi_k(s) \nabla) \phi_i(s) \right] \phi_j(s) ds,
\end{eqnarray*}
such that the overall mass and stiffness matrices can be written as
\begin{equation}
\mass(\bm\gamma) = \sum_{k = 1}^{J} \gamma_k  \mass^{k}, \quad {\rm and} \quad \stiff(\bm\kappa) = \kappa_k  \sum_{k = 1}^{J} \stiff^{k}, 
\end{equation}
where ${\bm\gamma} = [\gamma(s_1), \ldots, \gamma(s_J)]^\top$ and ${\bm\kappa} = [\kappa(s_1), \ldots, \kappa(s_J)]^\top$.
Here, the local mass and stiffness matrices, $\mass^{k}$ and $\stiff^{k}$, can be precomputed for a given set of finite element basis functions. 
Following the recursive definition given in \cite{lindgren11}, the precision matrix $\precision(\alpha, {\bm\gamma}, {\bm\kappa})$ parameterized by a scalar $\alpha$ and vectors ${\bm\gamma}$ and ${\bm\kappa}$ is
\begin{eqnarray}
\precision(\alpha, {\bm\gamma}, {\bm\kappa}) & = \stiff(\bm\kappa) + \mass(\bm\gamma), & \quad \alpha = 1, \nonumber \\
\precision(\alpha, {\bm\gamma}, {\bm\kappa}) & = \precision(\alpha-1, {\bm\gamma}, {\bm\kappa}) \mass(\bm\gamma)^{-1} \left( \stiff(\bm\kappa) + \mass(\bm\gamma) \right), & \quad \alpha > 1. 
\end{eqnarray}

In the present work, we will pre-select the order of the differential operator by choosing an $\alpha$ value, $\alpha = \hat{\alpha}$, where $\hat{\alpha}\in \{2, 3, 4\}$. A larger exponent, e.g., $\alpha > 4$, is not recommended as it will lead to a high computational cost in the following parameter estimation problem. 
For a fixed $\hat{\alpha}$, the parameters ${\bm\gamma}$ and ${\bm\kappa}$ can be estimated in a Bayesian manner. 
The likelihood function for this estimation problem takes the form
\begin{eqnarray}
& p\left(\mathbf{x}^{(1)},\ldots, \mathbf{x}^{(N)}  \vert  {\bm\gamma} , {\bm\kappa}\right)  = \nonumber \\ & \quad\sqrt{ (2\pi)^d \,|\precision(\hat{\alpha}, {\bm\gamma}, {\bm\kappa})| }  \exp \left( - \frac12 \sum_{i=1}^N (\mathbf{x}^{(i)}-\bm{\mu})^\top \precision(\hat{\alpha}, {\bm\gamma}, {\bm\kappa}) (\mathbf{x}^{(i)}-\bm{\mu}) \right).
\label{eq:like1}
\end{eqnarray} 
Choosing spatially varying parameters ${\bm\gamma}$ and ${\bm\kappa}$ provides the flexibility needed to model a non-stationary covariance structure. 
However, estimating ${\bm\gamma}$ and ${\bm\kappa}$ from a limited number of snapshots is itself an ill-posed inverse problem, so priors 
must be assigned to these parameters to remove the ill-posedness. 

To limit the degrees of freedom in the estimation, we prescribe the function $\gamma(s)$ to be a scalar, i.e., $\gamma(s) = \gamma$, and use only a spatially varying $\kappa(s)$ to control the nonstationarity of the resulting GP. 
We use the mass lumping technique to approximate the mass matrix rather than dealing with the computationally prohibitive inversion $\mass(\bm\gamma)^{-1}$.
For the case $\gamma(s) = \gamma$, the lumped mass matrix is given as
\[
\mass_L(\gamma)_{ij} = \gamma \left( \delta_{ij} \sum_{k = 1}^{J} \sum_{l = 1}^{J} \mass^{k}_{il} \right). 
\]
We use an exponential prior to enforce the positivity of $\gamma$. 
Furthermore, a smooth lognormal process is prescribed as the prior for $\kappa(s)$ to enforce the positive semi-definiteness of the stiffness matrix $\stiff(\bm\kappa)$. 
By defining ${\bm\nu} = \log({\bm\kappa})$, the posterior distribution of the Bayesian inference problem can be written as 
\begin{eqnarray}
& p\left( \gamma , {\bm\nu} \vert \mathbf{x}^{(1)},\ldots, \mathbf{x}^{(N)} \right)  \propto \nonumber \\ 
& \quad\sqrt{ (2\pi)^d \,|\precision(\hat{\alpha}, \gamma, {\bm\nu})| }  \exp \left( - \frac12 \sum_{i=1}^N (\mathbf{x}^{(i)}-\bm{\mu})^\top \precision (\hat{\alpha}, \gamma, {\bm\nu}) (\mathbf{x}^{(i)}-\bm{\mu}) \right) \label{eq:post} \\
& \quad \times  p(\gamma) \times  \exp \left( - \frac12 (\bm\nu-\bm{\nu}_0)^\top \precision_\nu (\bm\nu-\bm{\nu}_0) \right),\nonumber
\end{eqnarray} 
where $({\bm\nu}_0, \precision_\nu)$ define the mean and precision of the lognormal prior for $\bm\kappa$.
The precision matrix $\precision(\hat{\alpha}, \gamma, {\bm\nu})$ in (\ref{eq:post}) is given as 
\begin{eqnarray}
\precision(\hat{\alpha}, \gamma, {\bm\nu}) & = \stiff(\exp(\bm\nu)) + \mass_L(\gamma), & \quad \hat{\alpha} = 1, \nonumber \\
\precision(\hat{\alpha}, \gamma, {\bm\nu}) & = \precision(\hat{\alpha}-1, \gamma, {\bm\nu}) \mass_L(\gamma)^{-1} \left( \stiff(\exp(\bm\nu)) + \mass_L(\gamma) \right), & \quad \hat{\alpha} > 1. 
\end{eqnarray}
%

We can then obtain the {\it maximum a posteriori} estimate of the GMRF parameters $(\gamma, {\bm\nu})$ by maximizing the logarithm of the posterior density, which yields
\begin{eqnarray}
\{ \hat{\gamma} , \hat{\bm\nu} \} & = &   \mathrm{arg\,max}_{\gamma , {\bm\nu}} \log  p\left( \gamma , {\bm\nu} \vert \mathbf{x}^{(1)},\ldots, \mathbf{x}^{(N)} \right) \nonumber \\
 & = &   \mathrm{arg\,max}_{\gamma , {\bm\nu}} \left \{ \frac12 \log(|\precision(\hat{\alpha}, \gamma, {\bm\nu})| )  - \frac12 \sum_{i=1}^N (\mathbf{x}^{(i)}-\bm{\mu})^\top \precision (\hat{\alpha}, \gamma, {\bm\nu}) (\mathbf{x}^{(i)}-\bm{\mu}) \right . \nonumber \\
 && \left . - \frac12 (\bm\nu-\bm{\nu}_0)^\top \precision_\nu (\bm\nu-\bm{\nu}_0) + 
\log p(\hat{\alpha}) + \log p(\gamma)  \right \}.
\label{eq:mrf_post}
\end{eqnarray}
The optimization problem (\ref{eq:mrf_post}) is continuous, and the gradient and Hessian of the objective with respect to $\gamma$ and ${\bm\nu}$ can be analytically derived (not reported here for brevity); hence, gradient-based nonlinear programing tools can be used to obtain an optimum. 
In particular, we employ the subspace trust region method of \cite{coleman94, coleman96} with inexact Newton steps. 
%
%
Note that in the inexact Newton solve, we use Hessian-vector products rather than explicitly forming the full Hessian, to ensure that computational costs and memory requirements scale favorably with parameter dimension. 

Given the solution of (\ref{eq:mrf_post}), the low-dimensional subspace for our reduced filters can be computed from the eigendecomposition of the estimated state covariance $\precision(\hat{\alpha}, \hat{\gamma}, \hat{\bm\nu})^{-1}$ using the matrix-free methods discussed in the PCA case (Section \ref{sec:pca}).

\begin{remark}
A subtle issue in constructing the precision matrix is the choice of boundary condition for the SPDE (\ref{eq:SPDE}). 
For a GP specified with scalar $\gamma$ and $\kappa$, prescribing a Dirichlet boundary condition will enforce zero variability on the boundary, and prescribing a no flux boundary condition---which is a common choice in the literature---will roughly double the variability at the boundary compare to the variability in the interior. 
Clearly both choices will result a GP that has nonstationary behavior near the boundary, even though the GP with scalar $\gamma$ and $\kappa$ defined on an infinite domain should be stationary in theory. 
In our numerical examples, boundary conditions do not present any difficulties, since we use periodic boundary conditions that are inherited from the structure of the corresponding data assimilation problem. 
In a more general setting, we recommend to use a zero-flux boundary condition and to let the data determine the nonstationary correlation structure through the solution of (\ref{eq:mrf_post}). This way, artifacts created by the boundary condition can be potentially compensated for via the inhomogeneous $\kappa(s)$ field.

%
%

\end{remark}
}

\section{Connection to ROKF}
\label{sec:rokf}

The reduced-order Kalman filter (ROKF), developed in \cite{cane96} and discussed further in \cite{hoteit04} is, in principle, very similar to the dimension reduction approaches presented in this paper; like our approach, ROKF uses a fixed low-dimensional subspace to represent the state vector. Therefore, it is useful to discuss the differences between the techniques in more detail. 

To see the difference between the methods, it is instructive to compare what kind of priors (predictive distributions) the methods induce for the subspace coordinates. The prior mean is propagated in the same way, but the covariance is handled differently. In the ROKF, the prediction covariance in the reduced space, denoted here by $\mathbf{\Psi}_k^f$, is computed by evolving $\mathbf{\Psi}_{k-1}^a$ as follows:
\begin{eqnarray}
\mathbf{\Psi}_k^f & = & (\mathbf{P}_r^\top\mathbf{M}_k\mathbf{P}_r) \mathbf{\Psi}_{k-1}^a (\mathbf{P}_r^\top\mathbf{M}_k\mathbf{P}_r)^\top+\mathbf{P}_r^\top\mathbf{Q}_k\mathbf{P}_r \\ 
 & = & \mathbf{P}_r^\top \mathbf{C}_k^f \mathbf{P}_r,
\end{eqnarray}
where $\mathbf{C}_k^f=(\mathbf{M}_k\mathbf{P}_r) \mathbf{\Psi}_{k-1}^a(\mathbf{M}_k\mathbf{P}_r)^\top+\mathbf{Q}_k$ is the posterior covariance of the previous time step propagated forward with $\mathbf{M}_k$. This has two interpretations: (\textit{a}) projecting the forward dynamics $\mathbf{M}_k$ onto the subspace spanned by the columns of $\mathbf{P}_r$ and using it to propagate the reduced covariance forward, and (\textit{b}) propagating the posterior covariance of the previous time step with the full model and projecting the resulting prediction covariance onto the subspace.

On the other hand, the prediction covariance in our approach (see Section \ref{sec:dyn}) is 
\begin{equation}
\mathbf{\Psi}_k^f = \left( \mathbf{P}_r^\top (\mathbf{C}_k^f)^{-1} \mathbf{P}_r \right)^{-1}.
\end{equation}
That is, while ROKF projects the predicted \textit{covariance} matrix $\mathbf{C}_k^f$, we project the predicted \textit{precision} matrix $(\mathbf{C}_k^f)^{-1}$. This difference has a nice geometric interpretation; projecting the covariance matrix is equivalent to {\it marginalizing} the prior onto the subspace, whereas projecting the precision matrix amounts to taking the {\it conditional} of the prior in the subspace. 

\begin{figure}[h!]
\begin{center}
\includegraphics[width=0.8\textwidth]{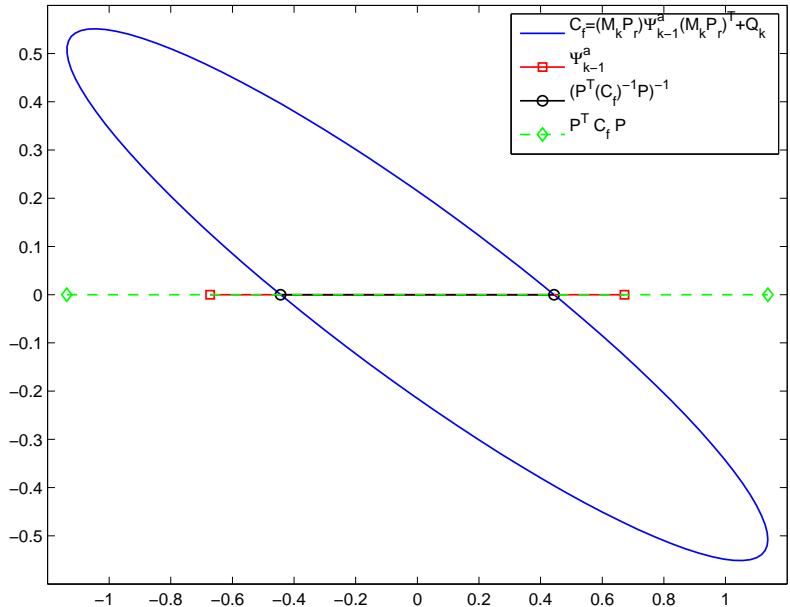}
\caption{Initial covariance in the subspace (red) propagated forward with the dynamical model (blue). Propagated covariance in the subspace using the ROKF (green) and our approach (black). The lines and ellipse contain 95\% of the probability mass of the associated 1-dimensional and 2-dimensional Gaussians.}
\label{fig:rokf_demo}
\end{center}
\end{figure}

To visually see the difference, consider the following simple example. Assume that the full dimension of the state space is $d=2$ and that the subspace (of dimension $r=1$) is the x-axis: $\mathbf{P}_r=[1, 0]^\top$. The model $\mathbf{M}_k$ is taken to be a $2 \times 2$ matrix with random entries sampled from $\mathrm{N}(0,1)$. We start with zero mean and covariance $\mathbf{\Psi}_{k-1}^a$ in the subspace (x-axis), and propagate it forward with both the ROKF and our approach. The results are shown in Figure~\ref{fig:rokf_demo}. One can clearly see that the prior induced by ROKF can be significantly different than the prior induced by our approach. Specifically, the prior induced by the ROKF is always wider than the conditional prior of our approach. It is clear that marginalizing the prior can yield posterior estimates which are \textit{outside the essential support of the prior.} 

\section{Numerical examples}

Here, we demonstrate the proposed dimension reduction algorithms with two synthetic filtering problems: a 240-dimensional version of the Lorenz model and a 1600-dimensional example using the quasi-geostrophic model.

{ As the reference methods, we use the standard extended Kalman filter (EKF) and the standard ensemble Kalman filter (EnKF); see \cite{evensen07}. The purpose of the experiments is to highlight some of the properties of the proposed approach, such as its behaviour with small sample sizes in ensemble filtering, rather than to draw conclusions about the performance of the approach relative to all the recent developments in the ensemble filtering literature. For this reason, and to keep the comparisons simple, we choose the well-known standard EnKF as the reference method instead of one of the many variants developed recently. A thorough performance comparison with all the recent developments in filtering methods is a challenging task (e.g., handling all the tuning issues of the various filters) and left for future research.}

For the EnKF, we implement a simple and widely used localization scheme, obtained by \textit{tapering} (setting the covariance between distant points to zero) the prediction covariance matrix using the 5th order piecewise rational function \cite{gaspari99}. The correlation cut-off length was chosen experimentally so that roughly optimal filter performance was obtained. 

\subsection{Example 1: Lorenz model II}
\label{sec:numex}

\subsubsection{Model description.}

As a small scale nonlinear example, we consider a generalized version of the Lorenz 96 model, the model II described in \cite{lor05}. The evolution model is given by an ODE system of $N$ equations, each defined as 
\begin{equation}
\frac{dX_n}{dt} = \sum_{j=-J}^{J} \sum_{i=-J}^{J} (-X_{n-2K-i}X_{n-K-j}+X_{n-K+j-i}X_{n+K+j})/K^2-X_n+F_n,
\label{mod}
\end{equation} 
where $n=1,\ldots,N$ and $K$ is a chosen \textit{odd} integer and $J=(K-1)/2$. The variables are periodic: $X_{-i}=X_{N-i}$ and $X_{N+i}=X_i$ for $i \geq 0$. With $K=1$, the system reduces to the standard Lorenz 96 model introduced in \cite{lor96}. 

\begin{figure}[h!]
\begin{center}
\scalebox{0.75}{\includegraphics{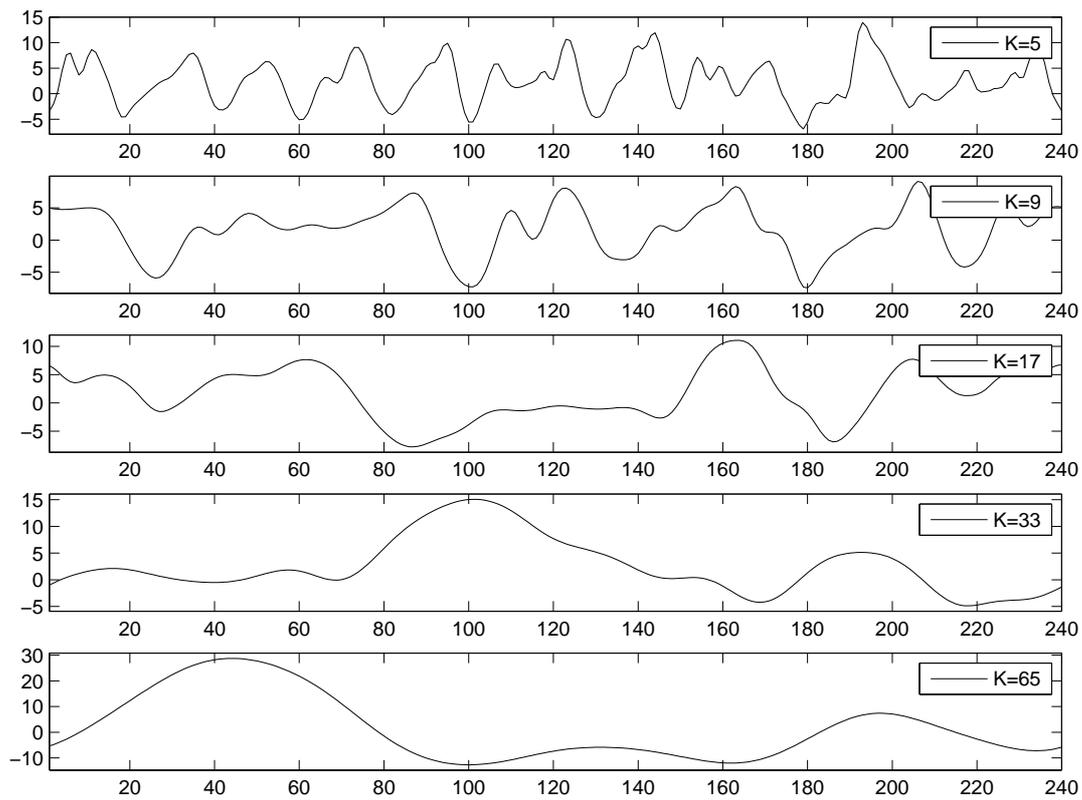}}
\caption{Solutions to the Lorenz model II with $N=240$ at one time step with different values for $K$. Larger $K$ yields smoother solutions.}
\label{fig:lorex}
\end{center}
\end{figure}

\begin{figure}[h!]
\begin{center}
\scalebox{0.7}{\includegraphics{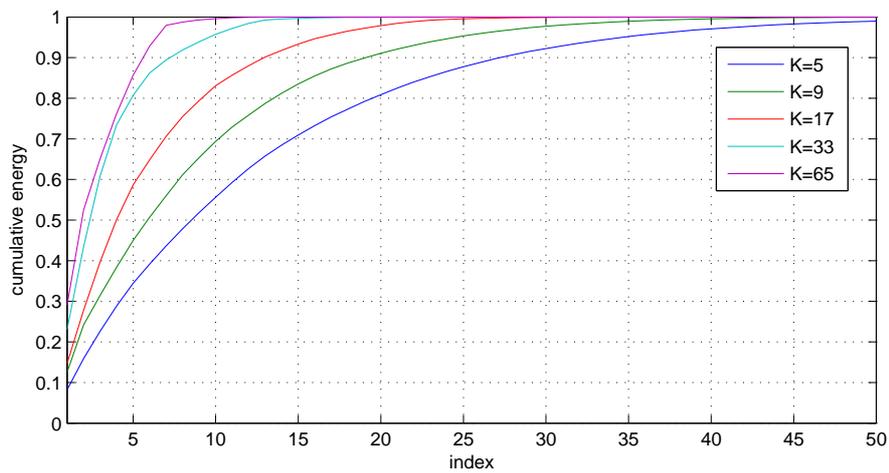}}
\caption{Cumulative energy of the empirical covariance matrix of the solution trajectories with varying $K$.}
\label{fig:energy}
\end{center}
\end{figure}

In our experiments we use a range of values for $K$, and choose the forcing $F_n$ so that the model attains chaotic behaviour (verified experimentally). In the prediction model used in the estimation, we use values $K=5,9,17,33,65$ and the corresponding forcing values $F_n=10,10,12,14,30$ for all $n$. Increasing $K$ introduces stronger spatial dependence between neighbouring variables, and yields spatially smoother solutions; example solutions with $N=240$ and varying $K$ are given in Figure \ref{fig:lorex}.
Controlling the smoothness allows us to demonstrate how the dimension reduction works in different cases: the smoother the unknown, the fewer basis vectors we need to describe it accurately. This is demonstrated in Figure~\ref{fig:energy}, where we plot the fraction of the energy of the empirical covariance matrix of the model trajectories as a function of the number of basis vectors used in the representation. More precisely, we plot $\sum_{i=1}^r \lambda_i/\sum_{i=1}^d \lambda_i$ as a function of $r$, where $\lambda_i$ is the $i$th largest eigenvalue of the empirical covariance matrix computed from model simulation output. One can see that with large $K$, most of the variability of the model state can be captured in a low-dimensional subspace, whereas with small $K$ more basis vectors are needed. 

\subsubsection{Experiment setup.}

In the experiments, we use values $K=5,9,17,33,65$ and the corresponding forcing values $F_n=10,10,12,14,30$ for all $n$. We generate data for the estimation by simulating the model and adding 1\% normally distributed random perturbations to the forcing values to introduce error into the prediction model. The observation frequency is $0.05$ time units (twice the time step used to solve the ODE) and every $10$th state variable is observed ($X_1,X_{11},\ldots,X_{231}$), which yields altogether 24 measurements per observation time step. Data is generated for $20$ time units, which yields altogether 400 observation times. { The ODE was solved using the 4th order Runge-Kutta method.}

The model error covariance matrix used in the experiments is simply $\mathbf{Q}_k=\beta \mathbf{I}$ for all $k$, where $\beta$ is chosen experimentally from the interval $\beta \in [0.01,0.3]$ so that roughly optimal tracking performance is obtained (RMS error between the estimates and the truth is minimised) for each filter.

The candidate subspaces were constructed by the PCA, GP, and GMRF techniques discussed in Section~\ref{sec:subspace}. For PCA, we used the trajectory obtained by simulating the model for 1200 steps (that is, the covariance was estimated using 1200 samples). GP and GMRF parameters were fitted using 32 snapshots of the model state. For the GP approach, the squared exponential covariance function was used, see equation (\ref{squared-exp}), and the obtained estimates for the variance and correlation length parameters were $\theta_1=6.42$ and $\theta_2=9.33$. {For the GMRF approach, we use an exponent of $\hat{\alpha} = 2$.}

\subsubsection{Results: EKF.}

\begin{figure}[h!]
\begin{center}
\includegraphics[width=\textwidth]{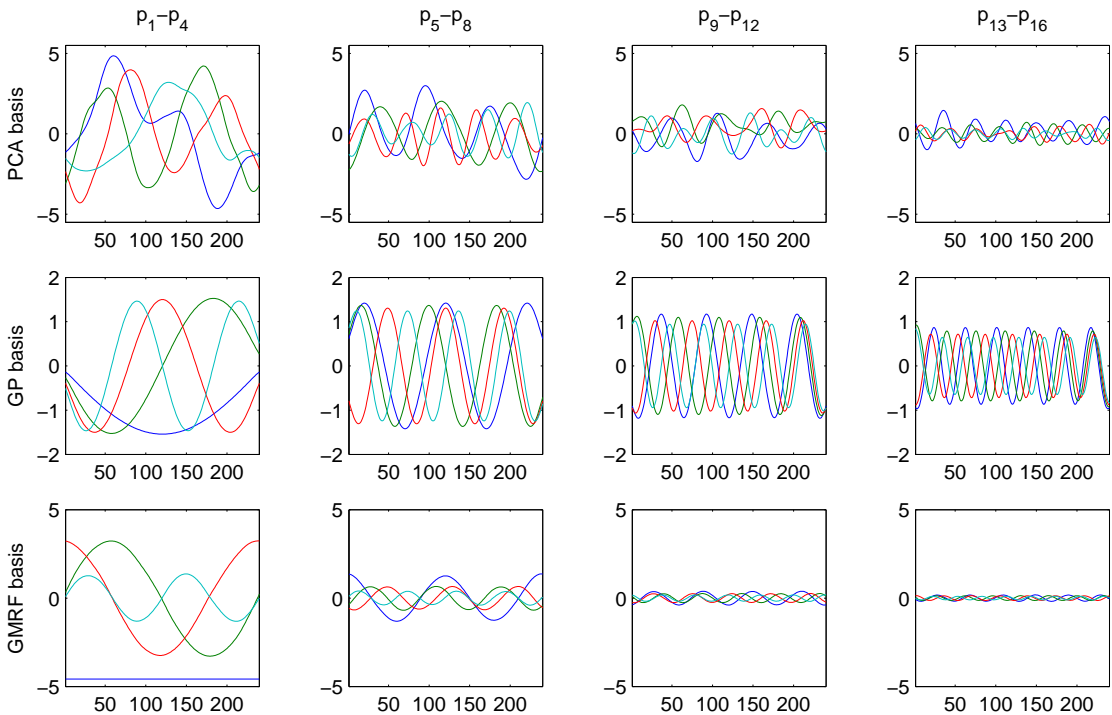}
\caption{The first 16 basis vectors obtained from PCA analysis and GP and GMRF fits.}
\label{fig:basis}
\end{center}
\end{figure}

First, we compare the reduced EKF described in Sections \ref{sec:redu} and \ref{sec:ekf} to the full EKF for the $K=33$ case, where the model state is spatially rather smooth and thus well described in a low-dimensional subspace; see Figures~\ref{fig:lorex}~and~\ref{fig:energy}. The first 16 basis vectors obtained via PCA, GP, and GMRF are given in Figure~\ref{fig:basis}; the first vectors represent large scale smooth features and the later ones describe finer scale features. 

\begin{figure}[h!]
\begin{center}
\scalebox{0.75}{\includegraphics{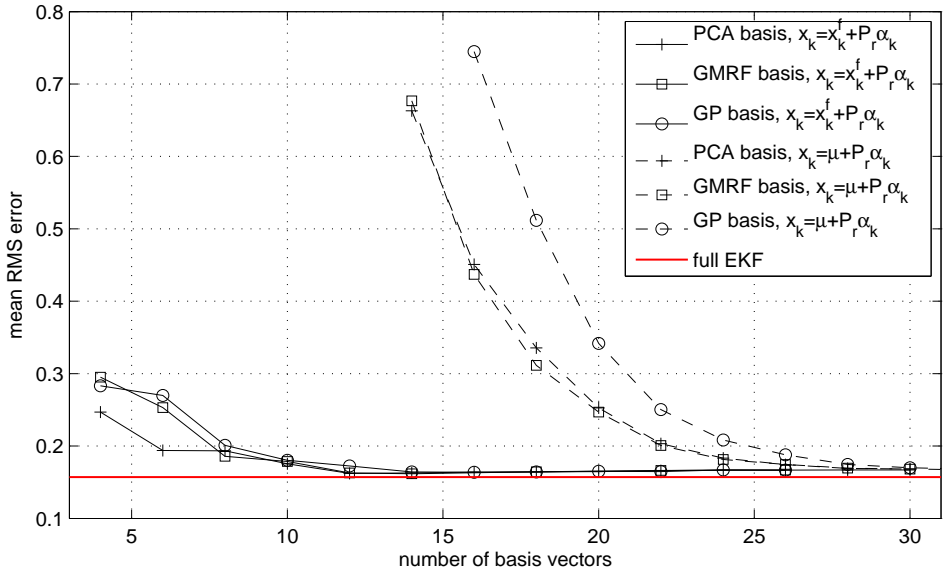}}
\caption{The average RMS error computed from steps 100--400 for the full EKF (red line) and for reduced EKF with increasing $r$ (black lines).}
\label{fig:lor_res1}
\end{center}
\end{figure}

All methods were started from an all-zero initial state, $\mathbf{x}_0=\mathbf{0}$ and identity covariance $\mathbf{C}_0^a=\mathbf{I}$. In Figure \ref{fig:lor_res1}, we compare the RMS error of the EKF and reduced EKF with varying numbers of basis vectors $r$, three ways for constructing the subspace (PCA, GP, and GMRF), and two ways to parameterize the unknown (solid lines: centered at the prior mean $\mathbf{x}_k=\mathbf{x}_k^f+\mathbf{P}_r\bm{\alpha}_k$, dashed lines: fixed offset $\mathbf{x}_k=\bm{\mu}+\mathbf{P}_r\bm{\alpha}_k$). As the fixed offset $\bm{\mu}$, we use the empirical mean of the simulated model trajectory for PCA and a constant $\bm{\mu}$ for GP and GMRF. 

We observe that centering the parameterization at the prior mean improves the results dramatically compared to fixed mean. With small $r$, the PCA basis works slightly better than GP and GMRF. With the fixed mean parameterization, PCA and GMRF work roughly equally well, and GP a little worse. But all in all, the three different ways of constructing the subspace all yield similar results.

In this example, we are able to obtain a reasonably accurate filter even with $r=4$ (!), whereas the fixed offset parameterization requires roughly $r=20$ for similar accuracy. Thus, we are able to reduce the dimension and the computational complexity almost two orders of magnitude compared to the full state dimension $N=240$. 

\begin{figure}[h!]
\begin{center}
\includegraphics[width=\textwidth]{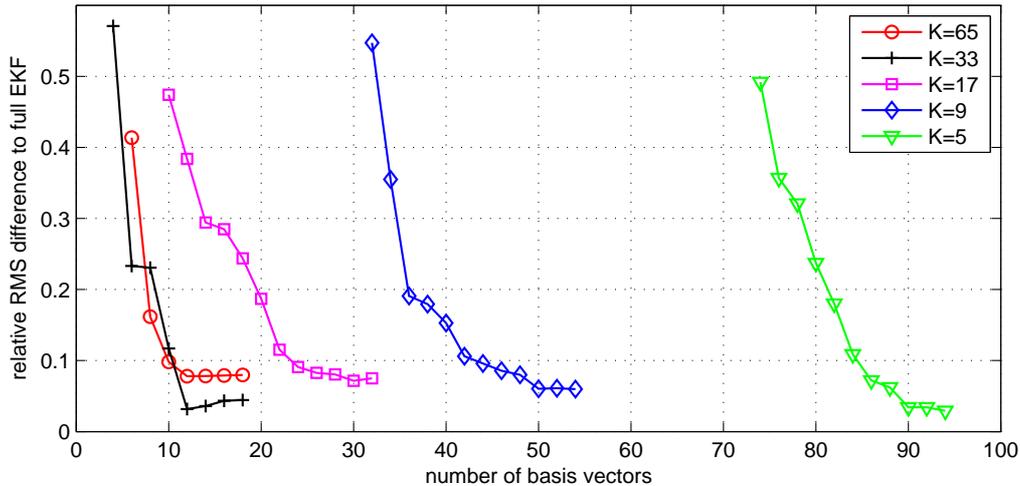}
\caption{Relative difference of the mean RMS error obtained by reduced EKF compared to full EKF with various values for $K$ and $r$.
}
\label{fig:lor_res_allk}
\end{center}
\end{figure}

A summary of the results for all cases $K=5,9,17,33,65$ using the PCA basis is given in Figure \ref{fig:lor_res_allk}. We plot the relative difference between the RMS error obtained with the full EKF and with the reduced EKF for all cases with varying $r$, using only the parameterization $\mathbf{x}_k=\mathbf{x}_k^f+\mathbf{P}_r\bm{\alpha}_k$. We observe that with high $K$ (when the trajectories are smooth), a small $r$ is enough to capture the system; for instance, with $K=65$ and $K=32$, only roughly $r=8$ vectors are needed to get an accurate filter. With small $K$, however, the system contains more fine scale features and larger $r$ is needed to obtain good filtering performance; for example, $K=5$ requires roughly $r=85$ for similar accuracy. This example illustrates how the efficiency of the proposed approach depends on the smoothness properties of the system. 

\subsubsection{Results: EnKF.}

\begin{figure}[h!]
\begin{center}
\scalebox{0.85}{\includegraphics{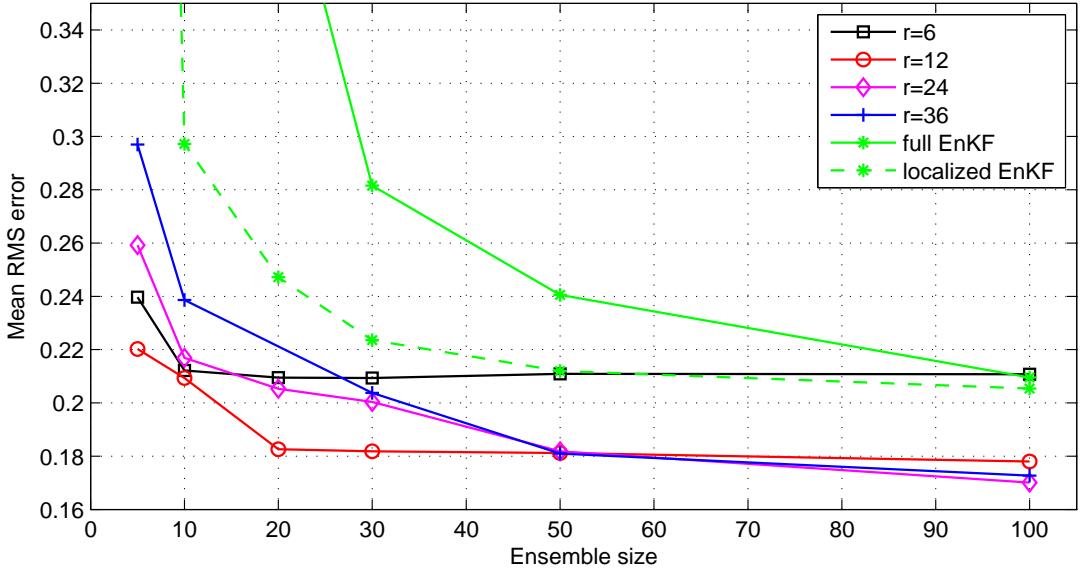}}
\caption{Mean RMS errors for full EnKF (with and without localization) and reduced EnKF with varying ensemble size and number of basis vectors $r$ used. For $N_{\mathrm{ens}}<20$, the EnKF results are cropped off because of filter divergence.}
\label{fig:lor_res3}
\end{center}
\end{figure}

Here, we compare the reduced EnKF described in Section \ref{sec:enkf} to the standard EnKF. 
The results for the $K=33$ case with varying $r$ are given in Figure \ref{fig:lor_res3} for different ensemble sizes. We observe that the reduced EnKF works much better than the standard EnKFs with small ensemble sizes; with the reduced EnKF, we are able to obtain a convergent filter with an ensemble size as small as $N_{\mathrm{ens}}=5$, whereas the standard EnKF has problems converging with $N_{\mathrm{ens}} \leq 20$. For example, the reduced EnKF with $r=12$ and $N_{\mathrm{ens}}=5$ yields similar performance as full EnKF with $N_{\mathrm{ens}}=100$. Again, with sufficiently high $N_{\mathrm{ens}}$ (here $N_{\mathrm{ens}} \geq 100$) the EnKF performance starts to catch up. 

Figure~\ref{fig:lor_res3} also illustrates the interesting connection between $r$ (number of basis vectors used) and $N_{\mathrm{ens}}$. With smaller $r$, the performance that can be obtained is poorer, but, on the other hand, a smaller $N_{\mathrm{ens}}$ is needed to achieve that performance. With larger $r$, better performance can be obtained, but only if $N_{\mathrm{ens}}$ is set high enough. That is, for each $N_{\mathrm{ens}}$, there seems to be an $r$ that provides an optimal compromise between representation error (due to small $r$) and sampling error (due to small $N_{\mathrm{ens}}$). 

We conclude that in ensemble filtering, the dimension reduction approach, when feasible, can offer a way to develop a reasonably accurate filter with fewer ensemble members. Restricting the inference to a subspace can reduce the need for sample covariance matrix regularization via localization techniques, which otherwise are needed when high-dimensional filtering problems are solved with small ensemble sizes \cite{anderson03,ott04}.

\subsection{Example 2: the two-layer quasi-geostrophic model}
\label{sec:numex2}

Next, we test the subspace filtering algorithms using the two-layer quasi-geostrophic model (QG model) \cite{Fandry1984}, which is often used as a benchmark system for data assimilation studies for numerical weather prediction (NWP). The model provides a reasonably good analogue of large-scale mid-latitude chaotic dynamics, while being relatively cheap computationally \cite{fisher11}. Next, we briefly describe the model equations and our estimation setup. For more details about the model as we use it, refer to \cite{fisher11}. 

\subsubsection{Model description.}

\begin{figure}[h!]
\begin{center}
\scalebox{1.1}{\includegraphics{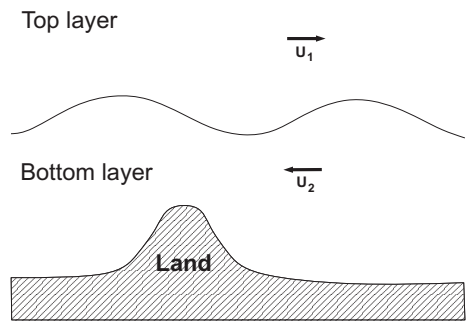}}
\caption{Geometrical layout of the two-layer quasi-geostrophic model.}
\label{qg_geometry}
\end{center}
\end{figure}

The two-layer quasi-geostrophic model simulates atmospheric flow for the geo\-stro\-phic (slow) wind motions. The geometrical domain of the model is specified by a cylindrical surface vertically divided into two ``atmospheric'' layers. The model also accounts for an orographic component that defines the surface irregularities affecting the bottom layer of the model. The latitudinal boundary conditions are periodic, whereas the values on the top and the bottom of the cylindrical domain are user-supplied constant values. The geometrical layout of the two-layer QG model mapped onto a plane is illustrated in Figure~\ref{qg_geometry}. In the figure, parameters $U_1$ and $U_2$ denote mean zonal flows in the top and the bottom atmospheric layers, respectively. The model formulation we use is dimensionless, where the non-dimensionalization is defined by the length scale $L$, velocity scale $U$, and the layer depths $D_1$ and $D_2$. 

The model operates with variables called {\it potential vorticity} and {\it stream function}, where the latter is analogous to pressure. The model is formulated as a coupled system of PDEs (\ref{qg_conservation}) describing a conservation law for potential vorticity. The conservation law is given as
\begin{equation}
\frac{D_1q_1}{Dt}=0,\ \frac{D_2q_2}{Dt}=0,
\label{qg_conservation}
\end{equation}
where $D_i$ denotes the substantial derivatives for latitudinal wind $u_i$ and longitudinal wind $v_i$, defined as $\frac{D_i\cdot}{Dt}=\frac{\partial\cdot}{\partial t}+u_i\frac{\partial\cdot}{\partial x}+v_i\frac{\partial\cdot}{\partial y}$; $q_i$ denote the potential vorticity functions; index $i$ specifies the top atmospheric layer ($i=1$) and the bottom layer ($i=2$). Interaction between the layers, as well as relation between the potential vorticity $q_i$ and the stream function $\psi_i$, is modeled by the following system of PDEs:
\begin{equation}
\label{pv_eq_1}
q_1=\nabla^2\psi_1-F_1\left(\psi_1-\psi_2\right)+\beta y,
\end{equation} 
\begin{equation}
\label{pv_eq_2}
q_2=\nabla^2\psi_2-F_2\left(\psi_2-\psi_1\right)+\beta y+R_s.
\end{equation}  
Here $R_s$ and $\beta$ denote dimensionless orography component and the northward gradient of the Coriolis parameter, which we hereafter denote as $f_0$. The relations between the physical attributes and dimensionless parameters that appear in (\ref{pv_eq_1})--(\ref{pv_eq_2}) are as follows:
\begin{eqnarray*}
&F_1=\frac{f_0^2L^2}{\acute{g}D_1},\ F_2=\frac{f_0^2L^2}{\acute{g}D_2},\ \acute{g}=g\frac{\Delta\theta}{\bar{\theta}},\\
&R_s=\frac{S\left(x,y\right)}{\eta D_2},\ \beta=\beta_0\frac{L}{U},
\end{eqnarray*}  
where $\Delta\theta$ defines the potential temperature change across the layer interface, $\bar{\theta}$ is the mean potential temperature, $g$ is acceleration of gravity, $\eta=\frac{U}{f_0L}$ is the Rossby number associated with the defined system, and $S(x,y)$ and $\beta_0$ are dimensional representations of $R_s(x,y)$ and $\beta$, respectively. 

The system of (\ref{qg_conservation})--(\ref{pv_eq_2}) defines the two-layer quasi-geostrophic model. The state of the model, and thus the target of estimation, is the stream function $\psi_i$. For the numerical solution of the system, we consider potential vorticity functions $q_1$ and $q_2$ to be known, and invert the spatial equations (\ref{pv_eq_1}) and (\ref{pv_eq_2}) for $\psi_i$. More precisely, we apply $\nabla^2$ to equation (\ref{pv_eq_1}) and subtract $F_1$ times (\ref{pv_eq_2}) and $F_2$ times (\ref{pv_eq_1}) from the result, which yields the following equation:
\begin{eqnarray}
&\nabla^2\left[\nabla^2\psi_1\right ]-\left(F_1+F_2\right)\left[\nabla^2\psi_1\right ]=\nonumber\\
&\nabla^2 q_1-F_2\left(q_1-\beta y\right)-F_1\left(q_2-\beta y-R_s\right).
\label{qg_helmholtz_eq}
\end{eqnarray}
Equation (\ref{qg_helmholtz_eq}) can be treated as a non-homogeneous Helmholtz equation with negative parameter $-\left(F_1+F_2\right)$ and unknown $\nabla^2\psi_1$. Once $\nabla^2\psi_1$ is solved, the stream function for the top atmospheric layer is determined by a Poisson equation. The stream function for the bottom layer can be found by plugging the obtained value for $\psi_1$ into (\ref{pv_eq_1}), (\ref{pv_eq_2}) and solving the equations for $\psi_2$. The potential vorticity functions $q_i$ are evolved over the time by a numerical advection procedure which models the conservation equations (\ref{qg_conservation}).

\subsubsection{Experiment setup.}

\begin{figure}[h!]
\begin{center}
\includegraphics[width=\textwidth]{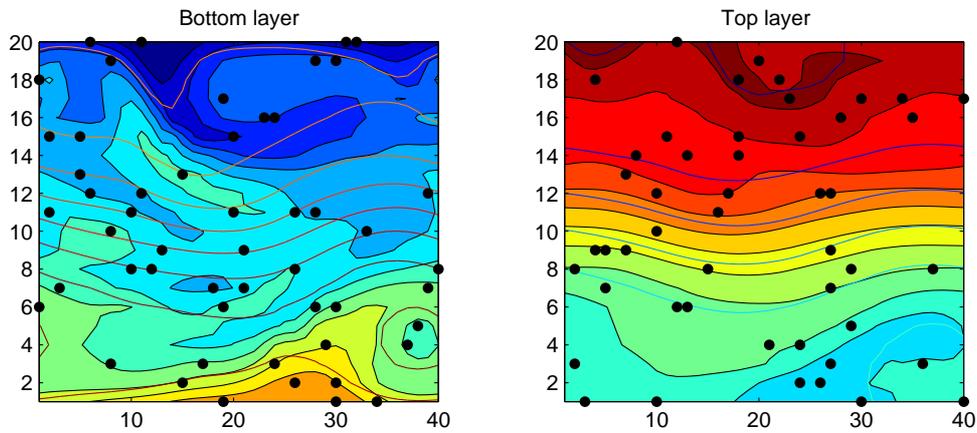}
\caption{A snapshot of the true state. The measurement locations are given as black dots. Contour lines in the background represent stream function (target of estimation) and the filled contours represent the potential vorticity.}
\label{fig:qg_setup}
\end{center}
\end{figure}

We run the QG model with $20 \times 40$ grid in each layer, and the dimension of the state vector is thus 1600. To generate data, we run the model with 1 hour time step using layer depths $D_1=6000$ and $D_2=4000$. Data is generated at every 6th step (filter step is thus 6 hours) by adding random noise for 100 randomly chosen grid points with standard deviation $\sigma=2.5 \cdot 10^{-3}$. For the estimation, bias is introduced to the forward model by using wrong layer depths, $\tilde{D}_1=5500$ and $\tilde{D}_2=4500$. The model error covariance matrix was $\mathbf{Q}=\beta \mathbf{I}$, where $\beta=10^{-4}$ was chosen experimentally so that roughly optimal EKF performance was obtained (here, the tracking performance was quite insensitive to $\beta$). A snapshot of the model simulation at one time step and the measurement locations are illustrated in Figure~\ref{fig:qg_setup}. 

\subsubsection{Results: EKF.}

\begin{figure}[h!]
\begin{center}
\includegraphics[width=\textwidth]{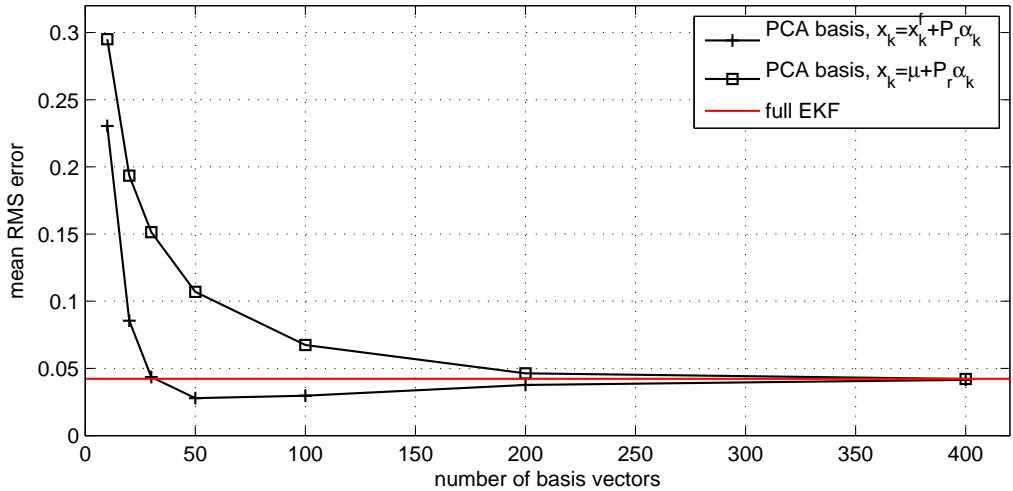}
\caption{Mean RMS errors of the full EKF and the reduced EKF with two parameterizations as a function of the number of basis vectors used.}
\label{qg_res}
\end{center}
\end{figure}

In Figure \ref{qg_res} we compare the mean RMS errors (computed over 400 filter steps) of the full EKF and the reduced EKF with two different parameterizations and different numbers of basis vectors $r$ used. Only the PCA basis is considered here. We observe, as in the other example, that the parameterization centered at the predicted mean works much better, especially with a small $r$. We are able to obtain reasonably accurate filtering results using only $r=20$ basis vectors, which reduces the CPU time (and memory requirements) by almost two orders of magnitude compared to full EKF. Interestingly, with a sufficiently high $r$, the average RMS error is actually \textit{lower} than with the full EKF. This can be explained by the additional prior information brought into the problem by restricting the inference onto a subspace. When $r$ approaches the full dimension of the problem $d$, all methods agree. 

\subsubsection{Results: EnKF.}

\begin{figure}[h!]
\begin{center}
\includegraphics[width=\textwidth]{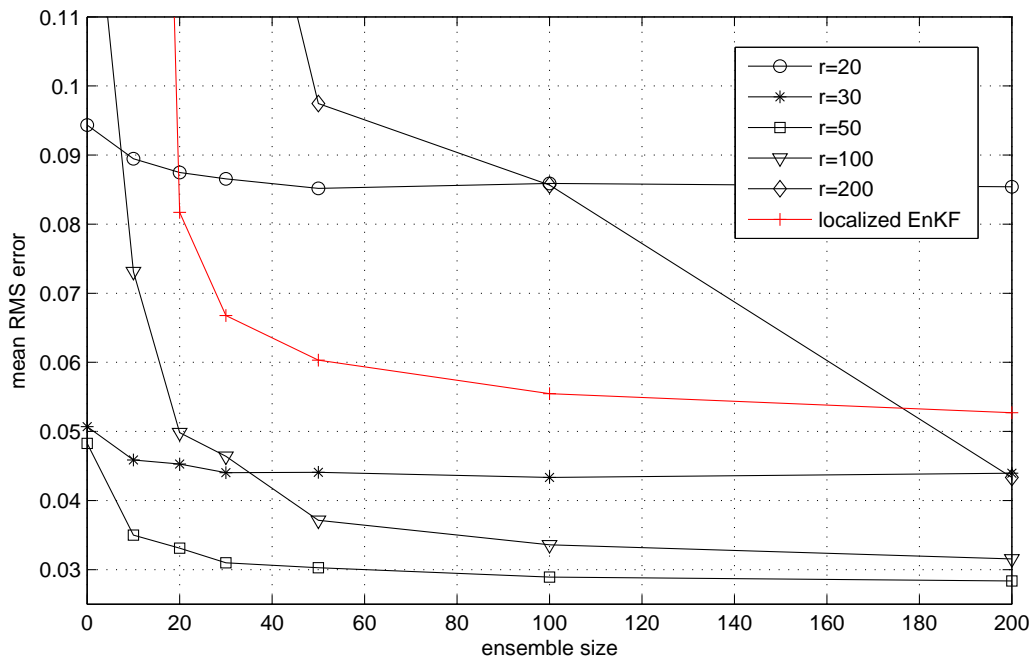}
\caption{Mean RMS errors as a function of ensemble size with varying number of basis vectors $r$ used. Ensemble size 0 means that a fixed prior covariance was used, without error propagation:  $\mathbf{C}_k^f=\mathbf{X}_k\mathbf{X}_k^\top+\mathbf{Q}_k=\mathbf{Q}_k$.}
\label{qg_res_enkf}
\end{center}
\end{figure}

Next, we run the reduced EnKF using the parameterization centered at the predicted mean, $\mathbf{x}_k=\mathbf{x}_k^f+\mathbf{P}_r\bm{\alpha}_k$. In Figure \ref{qg_res_enkf}, we plot the mean RMS error computed over 400 filter steps for varying ensemble sizes and varying number of basis vectors $r$. The results qualitatively follow the same pattern as with the Lorenz model in Section \ref{sec:numex}; with large $r$, a larger ensemble size is needed to get an accurate filter, and with small $r$, a smaller ensemble size is sufficient to get close to the optimal performance that is achievable with that $r$. For instance, with $r=50$, ensemble size $N_{\mathrm{ens}}=10$ yields better tracking performance than $r=200$ with $N_{\mathrm{ens}}=200$. 

The results are much better than what could be obtained with the standard EnKF; here, $N_{\mathrm{ens}}>200$ would be required even to get the standard EnKF to converge; see the results of \cite{solonen14}. Localization methods dramatically improve EnKF performance; for comparison, in Figure \ref{qg_res_enkf} we show the results for a simple localization, where we taper the prediction covariance matrix again using the 5th order piecewise rational function \cite{gaspari99}, experimentally tuning the cutoff length in the localization to achieve roughly optimal performance. However, the subspace algorithms still yield better results, as in the Lorenz example. Restricting the filtering onto a subspace regularizes the problem enough so that the need for localization is diminished.

Figure~\ref{qg_res_enkf} contains some results with ensemble size $N_{\mathrm{ens}}=0$. Here, zero ensemble size means that no samples were used to propagate the uncertainty (only the posterior mean was propagated), and the prediction covariance was taken to be the model error directly: $\mathbf{C}_k^f=\mathbf{X}_k\mathbf{X}_k^\top+\mathbf{Q}_k=\mathbf{Q}_k$. For small $r$, this simple 3D-Var type of strategy with a fixed prior was enough to get a convergent filter. When the uncertainty propagation via samples was added and the sample size was increased, the filter accuracy was improved, as expected. The larger the value of $r$, the more crucial the uncertainty propagation. This behavior can be explained by the fact that restricting the inference onto a subspace already heavily regularizes the problem, and thus propagating the covariance accurately is less important. For instance, using a small $r$ restricts the inference to a subspace spanned by spatially smooth basis vectors. In the full space algorithms, such smoothness information would be obtained by propagating the covariance forward in time. In the subspace method with small $r$, non-smooth directions are explicitly removed from the estimation problem, and even a simple, fixed prior can yield reasonably accurate results. Note also that here the number of observations is 100, which is, in many cases, larger than the dimension of the subspace, making the estimation problems numerically well posed and the role of the prior less important. 

\section{Discussion and conclusions}
\label{sec:conc}

In this paper, we presented an effective and simple-to-implement dimension reduction strategy for solving non-stationary inverse problems in a Bayesian framework.
By identifying a global reduced subspace that captures the essential features of the state vectors, we provided a new subspace-constrained Bayesian estimation technique for reducing the computational cost of filtering algorithms. 

Our approach is first applied to the Kalman filter for linear Gaussian models, and then generalized to nonlinear problems via the extended and ensemble Kalman filters.
In the Kalman filter and extended Kalman filter cases, the computational savings of our subspace-constrained technique is due to two sources: (\textit{a}) the number of forward model simulations required in each prediction step is only equal to the reduced subspace dimensions, as the error is only propagated along the coordinates of subspace basis; and (\textit{b}) the update step can be formulated efficiently on the subspace coordinates. 
This way we also avoid handling matrices in the full dimension of the state space.
In the ensemble version, computational savings stem from the fact that when the inference is constrained into a low-dimensional subspace, fewer ensemble members are needed for covariance estimation compared to the full space approach. 
Also, the need for covariance localization techniques to regularize the predicted covariance is diminished.

Two approaches for constructing the reduced subspace are discussed. 
The first idea---widely used in model reduction community---is to obtain snapshots of typical model states (e.g., by performing a sufficiently long free model simulation) and to compute the leading eigenvectors of the resulting empirical state covariance matrix.
The second idea is to infer the state covariance matrix from a limited number of snapshots using a Gaussian process hypothesis.
This choice ``fills in'' the missing information about model states (due to a limited number of snapshots) using the correlation structure encoded in the particular choice of GP. 
We discussed GP constructions using either stationary kernels that directly specify the covariance matrix, or non-stationary differential operators that correspond to sparse precision matrices.
The GP construction also opens the door to other possible state covariance reconstruction approaches; for instance, one could infer the state covariance from previous data sets. 
We will investigate this extension in future work.

We demonstrated the performance of our approach using two numerical examples. 
The first one is a 240-dimensional Lorenz system, where the smoothness of the model states can be controlled with a tuning parameter. 
This rather low-dimensional example is used to demonstrate the performance of our dimension reduction approach in various regimes. 
For smooth settings, the dimension can be reduced dramatically (to less than 10) while still obtaining filtering accuracy---for both extended Kalman filtering and ensemble filtering---comparable to the full space algorithms. 
%
%
On the other hand, for non-smooth settings with ``rough'' features, the level of dimension reduction that maintains filter accuracy becomes less dramatic. 
The second example is a 1600-dimensional two-layer quasi-geostrophic model, where the state dimension can be a reduced to about 30 without losing filtering accuracy for both the extended and ensemble filters, compared to their full space counterparts. 
A two order of magnitude reduction in computing time is achieved for extended Kalman filtering in this case.
%
%
In ensemble filtering, our subspace approach yields accurate filtering results with smaller ensemble sizes than the standard EnKF with localization.

\section*{Acknowledgements}

We thank Alexander Bibov for providing the quasi-geostrophic model implementation. A.\ Solonen and J.\ Hakkarainen acknowledge support from the Academy of Finland (project numbers 284715 and 267442). T.\ Cui and Y.\ Marzouk also acknowledge support from the US Department of Energy, Office of Advanced Scientific Computing Research (ASCR) under grant number DE-SC0003908.


\end{document}